%% file: main.tex
\documentclass[acmsmall,manuscript]{acmart}
\pdfoutput=1
\usepackage{booktabs} 
\usepackage[utf8]{inputenc}
\usepackage[ruled]{algorithm2e} 
\usepackage{enumitem}
\usepackage{amsmath}
\usepackage{graphicx}
\usepackage{capt-of}

\usepackage{color}
\usepackage{colortbl}
\usepackage{soul}          
\usepackage[draft,inline,nomargin]{fixme}
\usepackage{multirow}
\usepackage{setspace}
\usepackage{epstopdf}
\usepackage{booktabs}
\usepackage{array}
\usepackage{url}
\input{glyphtounicode}
\usepackage[T1]{fontenc}
\usepackage{subcaption}
\usepackage{tabularx}
\usepackage{ltxtable}
\usepackage{blindtext}
\usepackage{graphicx}
\usepackage{wrapfig,lipsum,booktabs}
\usepackage{booktabs}
\usepackage{array}
\def\mybar[#1]#2{
  {\color{black}\rule[0.1ex]{#1mm}{5pt}} #2}
\usepackage{tcolorbox}

\SetAlFnt{\small}
\SetAlCapFnt{\small}
\SetAlCapNameFnt{\small}
\SetAlCapHSkip{0pt}
\IncMargin{-\parindent}

\newcommand\para[1]{{\vspace{5pt} \noindent {\bf #1}}}

\newcolumntype{L}[1]{>{\raggedright\let\newline\\\arraybackslash\hspace{0pt}}m{#1}}
\newcolumntype{C}[1]{>{\centering\let\newline\\\arraybackslash\hspace{0pt}}m{#1}}
\newcolumntype{R}[1]{>{\raggedleft\let\newline\\\arraybackslash\hspace{0pt}}m{#1}}

\begin{document}

\title{From User Surveys to Telemetry-Driven AI Agents: Exploring the Potential of Personalized Productivity Solutions}

\author{Subigya Nepal}
\orcid{0000-0002-4314-9505}
\affiliation{%
  \institution{Microsoft Research}
  \city{Cambridge}
  \state{MA}
  \country{USA}
}
\affiliation{%
  \institution{Stanford University}
  \city{Stanford}
  \state{CA}
  \postcode{94305}
  \country{USA}
}
\email{sknepal@stanford.edu}
\author{Javier Hernandez}
\orcid{0000-0001-9504-5217}
\affiliation{%
  \institution{Microsoft Research}
  \city{Cambridge}
  \state{MA}
  \country{USA}
}
\author{Talie Massachi}
\orcid{}
\affiliation{%
  \institution{Brown University}
  \city{Providence}
  \state{RI}
  \country{USA}
}
\author{Kael Rowan}
\orcid{}
\affiliation{%
  \institution{Microsoft Research}
  \city{Redmond}
  \state{WA}
  \country{USA}
}
\author{Judith Amores}
\orcid{}
\affiliation{%
  \institution{Microsoft Research}
  \city{Cambridge}
  \state{MA}
  \country{USA}
}
\author{Jina Suh}
\orcid{}
\affiliation{%
  \institution{Microsoft Research}
  \city{Redmond}
  \state{WA}
  \country{USA}
}
\author{Gonzalo Ramos}
\orcid{}
\affiliation{%
  \institution{Microsoft Research}
  \city{Redmond}
  \state{WA}
  \country{USA}
}
\author{Brian Houck}
\orcid{0000-0001-9730-7292}
\affiliation{%
  \institution{Microsoft}
  \city{Redmond}
  \state{WA}
  \country{USA}
}
\author{Shamsi T. Iqbal}
\orcid{}
\affiliation{%
  \institution{Microsoft Research}
  \city{Redmond}
  \state{WA}
  \country{USA}
}
\author{Mary Czerwinski}
\orcid{0000-0003-0881-401X}
\affiliation{%
  \institution{Microsoft Research}
  \city{Redmond}
  \state{WA}
  \country{USA}
}

\input{abstract}

\ccsdesc[500]{Human-centered computing~Empirical studies in HCI}
\ccsdesc[300]{Human-centered computing~User studies}
\ccsdesc[300]{Human-centered computing~Empirical studies in collaborative and social computing}
\ccsdesc[300]{Human-centered computing~Empirical studies in ubiquitous and mobile computing}

\keywords{Workplace, Information workers, Large Language Models, LLM, AI, Productivity, Telemetry, Conversational agents, Digital activity, Dashboard, Narratives, User research}
\maketitle
\vspace{-1em}
\begin{center}
\small\textcolor{red}{\textit{This paper has been accepted for publication at CSCW 2025 after undergoing peer-review; publication is currently underway.}}
\end{center}
\vspace{0.5em}
\renewcommand{\shortauthors}{Nepal and Hernandez et al.}

\input{1_introduction}
\input{3_related_work.tex}
\input{4_methodology.tex}
\input{5_corelation.tex}

\input{8_discussion.tex}
\input{9_conclusion.tex}

\bibliographystyle{ACM-Reference-Format}
\bibliography{paper_bib}
\input{appendix.tex}

\end{document}

%% file: abstract.tex
\begin{abstract}
Information workers increasingly struggle with productivity challenges in modern workplaces, facing difficulties in managing time and effectively utilizing workplace analytics data for behavioral improvement. Despite the availability of productivity metrics through enterprise tools, workers often fail to translate this data into actionable insights. We present a comprehensive, user-centric approach to address these challenges through AI-based productivity agents tailored to users' needs. Utilizing a two-phase method, we first conducted a survey with 363 participants, exploring various aspects of productivity, communication style, agent approach, personality traits, personalization, and privacy. Drawing on the survey insights, we developed a GPT-4 powered personalized productivity agent that utilizes telemetry data gathered via Viva Insights from information workers to provide tailored assistance. We compared its performance with alternative productivity-assistive tools, such as dashboard and narrative, in a study involving 40 participants. Our findings highlight the importance of user-centric design, adaptability, and the balance between personalization and privacy in AI-assisted productivity tools. By building on these insights, our work provides important guidance for developing more effective productivity solutions, ultimately leading to optimized efficiency and user experiences for information workers.

\end{abstract}

%% file: 1_introduction.tex
\section{Introduction}
\label{sec:intro}
Productivity plays a pivotal role in determining professional success. Organizations and individuals continuously strive to enhance their efficiency, improve time management, and optimize their work processes. The importance of effective productivity tools has become increasingly critical, particularly as studies reveal concerning trends: information workers spend an average of 28\% of their workweek on email, 45\% of professionals report that excessive meeting time reduces their overall productivity, and 85\% of leaders express difficulty in confidently assessing employee productivity in hybrid work settings~\cite{chui2012social, calendlyStateMeetings, microsoftHybridWork}. Despite having access to productivity data through workplace analytics tools, many workers struggle to translate this information into actionable improvements, highlighting the need for more effective, personalized productivity solutions.
The rapid advancements in Artificial Intelligence (AI), natural language processing, and machine learning technologies have introduced promising opportunities to address these challenges. Several studies have explored the potential of AI-based tools in various professional contexts, such as project management~\cite{Taboada2023, Dam2019}, team collaboration~\cite{Bansal2020OptimizingAF}, and decision-making~\cite{Jarrahi2018}. These studies have demonstrated AI's potential to not only automate routine tasks but also to provide intelligent insights and recommendations that can help users optimize their work processes and achieve their goals more efficiently.

To develop a truly effective productivity agent, it is crucial to understand user preferences and the factors that influence productivity. This paper presents a comprehensive two-phase study aimed at investigating user preferences for productivity agents and devising personalized solutions that cater to these preferences. In the first phase, we conducted a user survey to identify the ideal persona for a productivity agent by examining people's expectations and preferences concerning its features and behaviors, exploring various aspects such as productivity, communication style, agent approach, personality traits, personalization, and privacy. Our findings revealed users' specific needs and areas where a personalized productivity agent could offer the most significant benefits. We also found that striking a balance between personalization and privacy is essential, as well as the need for adaptability in tone and approach depending on the situation and user preferences.

In the second phase, we developed an AI-based productivity agent that leverages insights from the initial user survey, telemetry data from users and uses the state-of-the-art GPT-4 model~\cite{openai2023gpt4}. GPT-4 is a powerful Large Language Model (LLM)~\cite{naveed2023comprehensive} capable of generating human-like text and understanding context. At the time of this study, GPT-4 represented the state-of-the-art in language models, making it well-suited for providing personalized and contextually relevant productivity advice based on individual user data. We utilized the enterprise version of GPT-4, which functions purely as an inference engine, processing only high-level telemetry data without storing or using any information for model training. Note that our approach to productivity tools is grounded in personal informatics-- using productivity tracking systems to support self-reflection and behavior change~\cite{Kross2023}.
Personal informatics systems, which help individuals collect and reflect on personal data about their behaviors, activities, and habits, have been shown to support behavior change and goal achievement across various domains~\cite{nepal2022survey, 10.1145/1753326.1753409, 10.1145/2030112.2030166, Choe2014}. In the context of workplace productivity, personal informatics tools can help workers understand their work patterns, identify areas for improvement, and make informed decisions about changing their behaviors~\cite{10.1145/3134714, Ekhtiar2023, 10.1145/2858036.2858066, 10.1145/3544548.3581326, 10388088}. The effectiveness of these tools often depends on how they present information to users and support different stages of behavior change, from initial awareness to sustained action. We explored three distinct approaches to presenting productivity data. Each approach aligns with stages of the personal informatics model: a dashboard visualization supporting reflection through visual pattern exploration, a personalized narrative aiding integration and reflection by connecting data points into a coherent story, and an interactive chatbot agent facilitating action and reflection through data-driven guidance. To personalize the productivity agent, information workers participating in our study provided their workplace telemetry data, such as email and meeting habits, to the agent through a custom application that we developed. The agent used the shared high-level telemetry data to tailor its recommendations and insights based on each user's unique work habits and patterns, providing personalized assistance for a more effective and customized user experience. In a within-subjects study, we then evaluated the AI agent's performance and compared it with dashboard visualization and personalized narratives. This analysis allowed us to better understand user preferences and identify areas for improvement and refinement. 

Our objective in this comprehensive study was to understand the nuanced requirements and preferences of users regarding AI-driven productivity tools. As such, we focus on user preferences and subjective experiences rather than measuring productivity improvements quantitatively. To structure our investigation, we formulated specific research questions for each phase of the study. These questions guided our research methodology and helped us focus on key areas of user interest and technology application:

\begin{itemize}
\item \textit{\textbf{(RQ1)}: What are the ideal characteristics and capabilities desired by users in a productivity agent?} This question aimed to uncover the specific features, traits, and functionalities users seek in a productivity-enhancing AI tool, considering aspects like personalization, communication style, and privacy concerns.
\item \textit{\textbf{(RQ2)}: How does the AI productivity agent based on findings from RQ1 compare with traditional methods of data presentation (dashboard visualization and narrative summaries)?} This question seeks to evaluate the effectiveness of the AI productivity agent, designed in line with RQ1 findings, against more conventional data presentation methods. It focuses on user preference, engagement, and the perceived utility of each approach in enhancing workplace productivity.
\item \textit{\textbf{(RQ3)}: How effective is the AI productivity agent in understanding and addressing users' productivity challenges based on their telemetry data?} This question centers on the AI productivity agent's performance, particularly focusing on its capability to analyze and respond to individual user productivity needs using the gathered telemetry data.
\item \textit{\textbf{(RQ4)}: What are the perceived strengths, weaknesses, and potential areas for improvement in the AI productivity agent, visual dashboard, and AI-based personalized narrative?} This question seeks to gather insights on the comparative advantages and limitations of the three data presentation methods, as well as identify opportunities for enhancing their overall performance and user experience.
\end{itemize}

In addressing the research questions discussed above, we make several novel contributions:
\begin{itemize}
\item We provide the first systematic comparison of different personal informatics approaches (dashboard, narrative, and interactive agent) for workplace productivity enhancement using AI and workplace telemetry data.
\item Unlike previous studies that focused on general virtual assistants or limited aspects of productivity tools, our large-scale survey (N=363) provides comprehensive insights into user preferences specifically for AI-based productivity agents. For example, we found that the majority of the participants preferred the agent to provide proactive suggestions to enhance productivity and that they were comfortable sharing high-level telemetry data for enhancing the response provided by the productivity agent.

\item We demonstrate a novel application of LLMs in interpreting workplace telemetry data for personalized productivity guidance, showing how advanced language models can transform generic productivity metrics into personalized, actionable insights. We conducted a thorough evaluation of the AI agent's performance with N=40 participants, highlighting its ability to provide personalized insights and recommendations based on user telemetry data.
\item Our findings reveal important trade-offs between different approaches to presenting productivity data - for instance, while participants strongly preferred dashboards for initial data exploration (55\% ranking it first), they found the interactive agent more effective for obtaining personalized recommendations and engaging with their productivity data in depth (showing 3-5x longer engagement times). These empirically-derived insights contribute new knowledge to guide the design of future productivity tools.

\end{itemize}

As we continue to witness rapid advancements in AI, natural language processing, and machine learning technologies, there is a growing potential for further enhancing and refining productivity agents. The incorporation of LLMs in the development of our AI-based productivity agent demonstrates the power of language models in creating personalized, context-aware, and potentially efficient productivity-enhancing tools. Our findings underscore the importance of a user-centric approach to the design and development of AI-based productivity agents and highlight the potential of incorporating design principles, such as semantic zooming~\cite{Shneiderman} and progressive disclosure~\cite{progressivedisclosure}, to enhance their effectiveness and user experience. By building on the insights obtained from this study, future research can continue to refine and optimize productivity-enhancing tools and solutions, ultimately leading to improved efficiency and user-centric experiences for information workers. This paper makes meaningful contributions to the field of HCI and AI-assisted productivity, and we hope these can meaningfully guide practitioners in the development of effective, user-centric solutions in this domain.

%% file: 3_related_work.tex
\section{Related Work}
\label{sec:related_work}
The landscape of productivity tools has undergone a significant evolution in recent years. The field has progressed from rudimentary task managers and calendar applications to more sophisticated AI-driven virtual assistants, such as Microsoft Cortana (now rebranded as Microsoft Copilot), Siri, and Google Assistant. Although numerous studies have emphasized the effectiveness of these general-purpose virtual assistants in improving user efficiency and satisfaction in various contexts and domains~\cite{deBarcelosSilva2020, Kiseleva2016, Brill2019}, there remains a gap in exploring AI-driven productivity agents tailored specifically to individual needs. This gap becomes even more pronounced when considering the unique demands of information workers, who face challenges that often diverge from those of the broader workforce or consumers.

Information workers, who primarily engage in data-centric tasks, often require more specialized tools to address their unique professional demands. Although general-purpose virtual assistants offer some degree of support, they may not fully meet the distinct requirements of these professionals, such as scheduling meetings, summarizing research reviews, and preparing presentations or reports. In recent years, customized solutions—particularly conversational agents—have emerged as a promising alternative due to their adaptability to individual professional needs~\cite{diederich2022design, feng2020my}. However, most of the existing research on chat or conversational agents is focused on customer service~\cite{Feine2019MeasuringSE, cui2017superagent} or marketing and sales~\cite{Vaccaro2018, van2019chatbot}. Other common areas of study for agents include education~\cite{hayashi2013pedagogical} and health~\cite{bhirud2019literature, bell2019perceptions}. Despite this focus, some progress has been made in researching solutions for information workers. For example, various conversational agent-based approaches have been proposed to assist with scheduling and managing tasks~\cite{toxtli2018understanding, Gil2008}, managing distractions~\cite{Tseng2019OvercomingDD, Mark2018}, coordinating meetings~\cite{Cranshaw2017}, work reflections~\cite{Kocielnik2018} and even augmenting group decision-making~\cite{Shamekhi2018FaceVE}. Notably, the work by ~\citet{grover2020design} demonstrated the design of a chatbot-like agent, complemented by an emotionally expressive video avatar, that aimed to enhance productivity. Their productivity agent assisted N=40 users in scheduling focused tasks, monitoring distractions, and reflecting on their daily mood and goals through a standalone application. The authors reported that the agent led to an increase in scheduled time for focused tasks, and that the users felt more satisfied and productive with the agents. Similarly, ~\citet{kimani2019conversational} introduced Amber, a conversational agent designed to support information workers across a diverse range of tasks, such as scheduling, prioritizing tasks, task switching, providing break reminders, managing social media distractions, and reflecting on daily accomplishments. The researchers conducted a field study with 24 participants over six days and found promising results for the potential use of conversational agents in enhancing workplace productivity and well-being. A critical aspect of designing these agents is understanding user preferences. For example, \citet{kimani2019conversational} conducted an online survey with N=70 participants to investigate when information workers might engage with a conversational agent. The authors found that users needed help with reminders, scheduling tasks, and managing distractions. Another study by \citet{Ahire2022} explored the expectations of N=14 knowledge workers from such agents, with findings highlighting the importance of features related to scheduling, distraction management, and well-being.~\citet{Khaokaew2022} conducted a user study of N=40 workers over four weeks to identify user needs for not just conversational agents but for all kinds of digital assistants (DA). The authors reported that the participants imagined a DA that supported them by managing their (1) time, (2) tasks, and (3) information.  While these insights provide valuable information, there is still a pressing need for a more comprehensive exploration of the specific requirements of information workers. Previous studies have broadly addressed productivity tools without honing in on the unique challenges and needs of information workers. Our research aims to fill this gap by focusing specifically on how these professionals interact with and benefit from AI-driven productivity assistants. Note, in this study, `information workers' refer to professionals whose primary role involves acquiring, manipulating, and generating information~\cite{10.5555/299856.299859}.

The emergence of advanced language models such as GPT-3 and GPT-4, have significantly expanded the potential of AI-driven conversational agents. However, creating agents tailored to individual users' work habits and patterns remains relatively uncharted territory. Although earlier applications of conversational agents in AI productivity tools, as developed by ~\citet{grover2020design} and ~\citet{kimani2019conversational}, utilized telemetry or sensing data for customization, it has yet to be thoroughly explored, particularly with the latest LLM-based chat agents that offer a more flexible and powerful alternative to their predecessors due to their semantic understanding. Although our work builds upon these existing productivity agent research efforts, several key aspects differentiate our approach. For example, both ~\citet{grover2020design} and ~\citet{kimani2019conversational} relied on a custom emotion and context sensing platform, where the collected data was shared with the productivity agent through a secure database. In contrast, our approach leverages modern LLM and readily available workplace telemetry data through Viva Insights\footnote{Microsoft's workplace analytics tool that offers data-driven insights on productivity, well-being, and work habits.}, eliminating the need to install additional detection systems while maintaining user privacy. This builds on a growing body of work exploring passive sensing and telemetry data for productivity and well-being applications~\cite{nepal2022survey, 10.1145/3544549.3585626, 9566617, 10.1145/3414118, 10.1145/3637304, 10765083}, although our implementation uniquely combines such data with the semantic understanding capabilities of modern LLMs. Our focus is specifically on facilitating productivity improvement through self-reflection rather than real-time intervention. Recent studies have demonstrated how AI systems can effectively support self-reflection and behavior change across various domains~\cite{10.1145/3613904.3642081, 10.1145/3544549.3585614, 10.1145/3699761}. By utilizing GPT-4, our agent can provide more nuanced and contextually relevant responses compared to traditional rule-based conversational agents. We also explicitly compare our agent with alternative presentation modes (dashboard and narrative), providing insights into the relative effectiveness of different approaches. Furthermore, our initial user survey (N=363) provides a broader understanding of user preferences compared to previous studies (e.g., there was no initial user survey done in ~\citet{grover2020design}, N=70 in ~\citet{kimani2019conversational}). However, these prior systems had some advantages over our approach. For instance, ~\citet{grover2020design}'s use of an emotionally expressive avatar might create stronger engagement, while ~\citet{kimani2019conversational}'s focus on well-being through break reminders represents an important aspect that could be incorporated into future versions of our system.

Building on these distinctions, our research aims to conduct a comprehensive two-phase study that examines user preferences and incorporates telemetry data for personalization in AI-driven productivity agents. Our approach goes beyond the existing literature by offering an in-depth analysis of both user preferences and the affordances of telemetry data, which provides valuable insights into user behaviors and work patterns. This combination of large language models~(LLMs) with telemetry data enables more accurate and contextually relevant productivity insights than previously possible.
It should also be noted that the shift to powerful but proprietary LLMs introduces new challenges in privacy and security ~\cite{10.1145/3663384.3663401, 10.1145/3544548.3581376}. These include concerns about data retention, model training practices, and the handling of potentially sensitive workplace information. Current approaches to mitigating these risks include using LLMs purely for inference without data retention which might involve working with enterprise versions of LLMs that provide additional privacy guarantees, limiting the granularity of data processed by these models or using locally hosted LLMs~\cite{li2024personal, nepal2024contextual}. Despite these potential privacy challenges, the implementation of telemetry data with modern LLMs in the context of productivity enhancement remains relatively unexplored.
Our study's innovation lies in the use of advanced LLMs and how we leverage these models to interpret telemetry data, creating a personalized and dynamic productivity tool tailored to individual user needs. We have the position that our findings can pave the way for future research and contribute to the development of more sophisticated productivity tools tailored to the distinct challenges and needs of information workers.

%% file: 4_methodology.tex
\section{Methodology}
\label{sec:methodology}
\begin{figure*}
    \centering
    \includegraphics[width=0.9\textwidth]{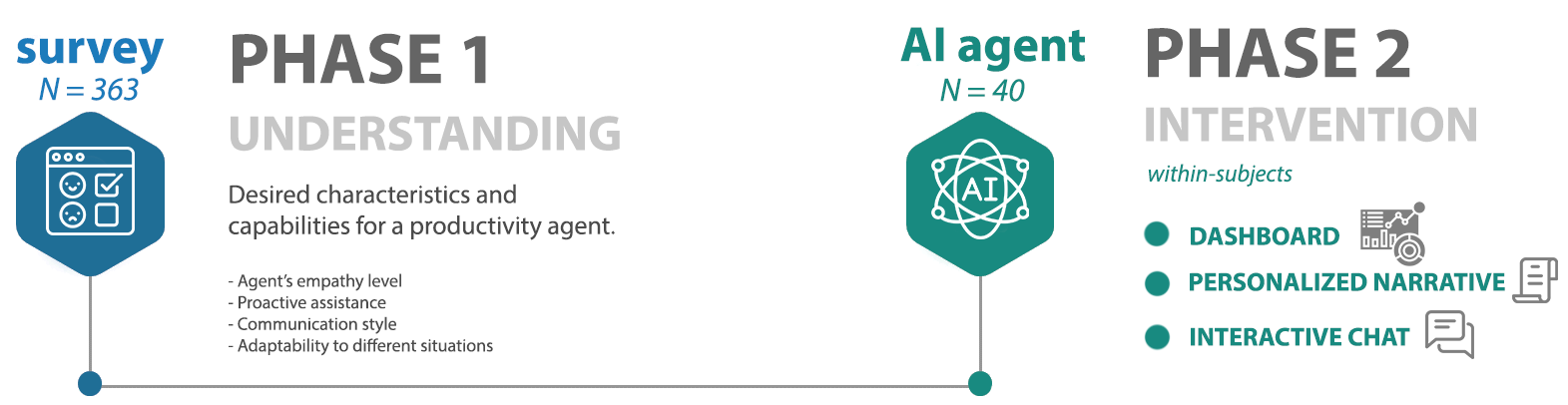}
    \caption{Two-Phase Study: In Phase 1, we explore desired productivity agent traits, while in Phase 2, we design and compare it with a visual dashboard and personalized narrative.}
    \label{fig:studyflow}
\end{figure*}

In this section, we present an overview of our study methodology, encompassing its design, data collection, and participant demographics. Our research methodology is divided into two distinct phases (as shown in \autoref{fig:studyflow}). The first phase involves conducting a user survey to gain insights into people's preferences regarding productivity agents. Utilizing the information gathered from the survey, the second phase focuses on designing a personalized productivity agent that caters to these user preferences. This two-phase approach ensures that our productivity agent designs are informed by users' needs and preferences, ultimately leading to more effective and user-centric solutions. All of the participants we enrolled are information workers working at a large multinational technology company based in the United States. The study was approved by the study institution's Internal Review Board (IRB).

\subsection{Understanding User Preferences for Productivity Agents}
The primary objective of our first survey was to identify the ideal persona for a productivity agent by examining people's expectations and preferences concerning its features and behaviors. We conducted a user survey with N=363 information workers, using this as an initial step to gather insights into their desired characteristics and capabilities for a productivity agent. The survey focused on various design aspects, including the agent's empathy level, proactive assistance, communication style, and adaptability to different situations. The survey we designed was carefully crafted based on prior work ~\cite{grover2020design,kimani2019conversational} and extensive interactions with experts in the field (i.e., AI professionals, organizational psychologists with expertise in workplace productivity and technology adoption, productivity engineers and expert HCI researchers). Drawing upon their knowledge and experience, we aimed to create a comprehensive and insightful questionnaire that would effectively capture users' preferences and concerns in relation to a productivity agent.  Refer to Appendix~\ref{sec:user_survey} for a comprehensive list of questions posed to participants. 

We distributed the survey randomly to employees within the company and received a total of 363 responses. The average duration participants spent on the survey was 17 minutes. Participation was anonymous and there was no compensation for this phase of the study.

\para{Demographics: } Appendix~\ref{sec:demographics_firstphase} presents the demographic information of our participants. The majority of the participants identified as male (67.2\%, N=244). In terms of age group, most participants were in the 36-45 (27.5\%, N=100), 46-55 (26.7\%, N=97), and 26-35 (26.4\%, N=96) age ranges. The majority of participants worked in software development/engineering roles (46.3\%, N=168), followed by product management (19.6\%, N=71).

\subsection{Productivity Solutions Design and Evaluation}
In the following phase of our study, we concentrated on developing an AI productivity agent specifically aimed at enhancing productivity. The design of the productivity agent was based on the insights obtained from the initial user survey, ensuring that the agent addresses the preferences and expectations of the participants. To further investigate the effectiveness of the proposed agent, we also asked participants to interact with two alternative methods of presenting telemetry data: a traditional dashboard visualization and an AI-based personalized narrative. 

\begin{figure}[ht!]
\centering
    \begin{subfigure}[b]{0.6\textwidth}
\hspace{1cm}\includegraphics[width=\textwidth]{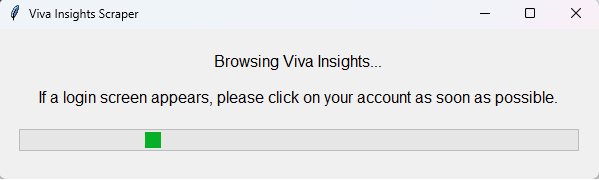}
\caption{Scraping in progress}
        \label{fig:scraping_first}
    \end{subfigure}%
    ~\newline
    \centering
    \begin{subfigure}[b]{0.7\textwidth}
        \includegraphics[width=\textwidth]{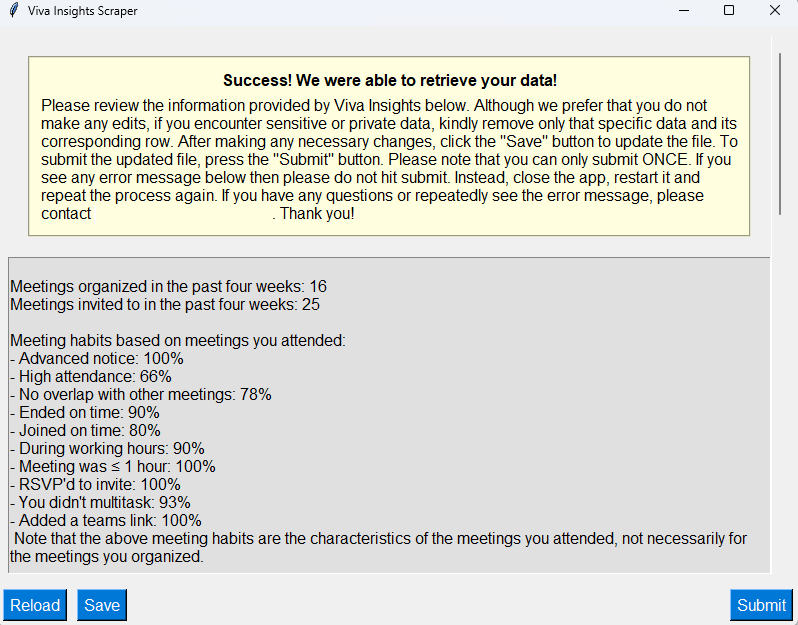}
        \caption{Final screen after successful scraping}
        \label{fig:scraping_last}
    \end{subfigure}
    \caption{Enhanced Telemetry Data Scraping: Figure (a) displays the user's initial view as our advanced scraping tool navigates through various Viva Insights pages to capture their source code. Once the source code is analyzed and the necessary data extracted, users will be presented with the final screen depicted in Figure (b). This concluding screen showcases the retrieved information, allowing users to review and modify the telemetry data prior to submission.}
    \label{fig:scraping}
\end{figure}

To personalize the experience for each individual, we decided to use the high-level telemetry data on participants' meetings, communication, collaboration network, and other work habits which are provided through Microsoft's Viva Insights\footnote{https://learn.microsoft.com/en-us/viva/insights/Use/Metric-definitions}. Among the different platforms, we selected Viva Insights as it is widely used in many companies and computes a wide variety of relevant features in the context of productivity. To extract the information, we built a custom dashboard scraper (see Figure~\ref{fig:scraping}) that allowed participants to automatically extract their Viva Insights data, review and modify it, and ultimately share it with the researchers. The extracted data was then used to create the three modes of presentation tailored to each participant's work patterns and habits. After participants willingly shared their telemetry data with us, they engaged with the three modes: viewing their data on a dashboard, reading a personalized narrative, and interacting with an AI productivity agent that had access to their telemetry data. They then assessed their experiences based on the ability to gain productivity insights or identify opportunities for productivity improvement by answering our follow-up survey. Note, the AI productivity agent primarily functioned as a `productivity guide,' offering recommendations and advice to improve users' productivity. This included suggesting strategies for time management, prioritizing tasks, and optimizing work routines. However, it is important to note that the agent did not directly perform tasks such as meeting summarization or documentation writing. While it provided guidance in these areas, the execution of tasks remained the user's responsibility. This distinction is crucial for understanding the agent's role in our study. In total, we had N=40 participants complete the second phase of our study. To diversify our sample, we reached out to individuals who had not participated in the first phase of the study. The average time for study completion was 30.15 minutes, and participants received a \$25 gift card as compensation. Our study exclusively makes use of OpenAI's~
\cite{openai} GPT-4 API. 

\begin{table*}[ht!]
\caption{Telemetry: List of telemetry data collected from the participants using our scraper tool. }
\small
\begin{tabular}{@{}lllll@{}}
\textbf{Category}                                                                                                                                          & \textbf{Telemetry Signal} \\ \bottomrule
{\cellcolor[HTML]{EFEFEF} Meeting} & total number of meetings organized and attended\\  & percentages of meetings organized that were one hour or shorter\\  & meeting invitations sent with more than 24 hours notice before the scheduled start time (i.e. advanced notice)\\ & meetings organized that included a Teams link for remote attendees\\
 & meetings organized or accepted during working hours\\
& online meetings (on Microsoft Teams) that ended within one minute of the scheduled end time \\  &  meetings organized or accepted that had a response rate over 50\% \\
 & online meetings (on Microsoft Teams) joined within five minutes of the scheduled start time\\
 & meetings that did not overlap with other meetings on the calendar \\
& meetings invited to and either accepted or declined (i.e., not having Tentative status) \\
 & meetings where the user did not read or send emails or chats (i.e., no multitasking)\\
{\cellcolor[HTML]{EFEFEF} Communication}  & total number of emails (sent and read)\\
& hourly breakdown of emails sent, emails read, and chats/calls made \\
{\cellcolor[HTML]{EFEFEF} Collaboration} & number of collaborators\\
& collaboration time within and outside working hours\\
{\cellcolor[HTML]{EFEFEF}Workplace Wellbeing }& total duration of focus time (i.e., uninterrupted work time) kept along with reserved focus time for next week \\
& number of days with disruptions (i.e. emails, chats, and meetings) after work hours\\
\bottomrule
\end{tabular}

  \label{tbl:telemetry}
\end{table*}

\para{Rationale for Comparative Analysis:}
The transition from Phase 1 to Phase 2 of our study was guided by the need to explore the practical application of user preferences identified in the initial survey. While Phase 1 established the preferred characteristics and functionalities of a productivity agent, Phase 2 aimed to contextualize these preferences within actual work settings. The introduction of a dashboard and narrative alongside the AI agent was a strategic decision to compare different modes of data presentation. This comparison was crucial to understanding how different formats could align with or diverge from the identified user preferences. The dashboard was considered for its traditional, structured approach to data visualization, while the narrative format was explored for its potential to provide a more engaging and story-like presentation of data. By comparing these with the AI agent, we sought to evaluate the effectiveness, user engagement, and preference alignment of each method, thereby offering a comprehensive view of how different data presentation strategies can cater to the diverse needs and preferences of information workers. This comparative analysis not only validates the effectiveness of our AI agent but also contributes to a broader understanding of user-centric data presentation in productivity tools.

\para{Telemetry Data Collection \& Challenges: } Our scraper tool employs Selenium~\cite{selenium}, a testing library, for browser automation to navigate and capture data from Viva Insights pages. The system's reliance on browser automation introduced several complexities: the scraping tool's functionality depended on OS compatibility and browser versions, requiring careful packaging of appropriate drivers (64-bit vs 32-bit) for different user environments. We encountered occasional timeout issues during processing, necessitating robust error handling.
Since these pages are dynamically generated, we opted for a more error-resilient approach by utilizing GPT-4 to extract information, rather than relying on traditional HTML tag-based scraping. However, this approach presented its own technical challenges, particularly regarding token limitations of the GPT-4 API. Our initial approach of passing entire HTML pages to GPT-4 frequently exceeded token limits. To address this, we implemented preprocessing steps using Python libraries to remove unnecessary HTML elements and select only relevant div tags before querying GPT-4. The streamlined source was then sent to GPT-4 with specific prompts to identify the desired information.
The integration of telemetry data with GPT-4 required careful prompt engineering to ensure consistent interpretation and reference. We structured the system prompt to include telemetry data in a standardized format, with clear delineation between different metrics (meetings, emails, focus time, etc.) to enable reliable data access. Since users had varying availability of telemetry data, we had to identify and focus on metrics that were commonly available across users. Our prompts were designed to explicitly exclude unavailable data points to prevent inconsistent or erroneous responses. To address these variations, we developed a template based on JSON that was used to dynamically complete the telemetry data portion of the system prompt, ensuring all necessary context was preserved regardless of data volume.
Please refer to Appendix~\ref{sec:prompts_extracting_data} and Appendix~\ref{sec:prompts_transfomring_data} for the comprehensive list of system prompts and user prompts we provided to GPT-4 for extracting and transforming the desired information. The telemetry information obtained from Viva Insights can be found in Table~\ref{tbl:telemetry}. It is important to note that Viva Insights calculates these insights based on the data from the prior month. Despite these various technical challenges, we successfully implemented web systems for both the GPT-4 agent and dashboard that provided seamless user experiences without visible errors, though with some latency in response times.

\begin{figure}[ht!]
 \begin{subfigure}[b]{0.48\textwidth}
    \includegraphics[width=\textwidth]{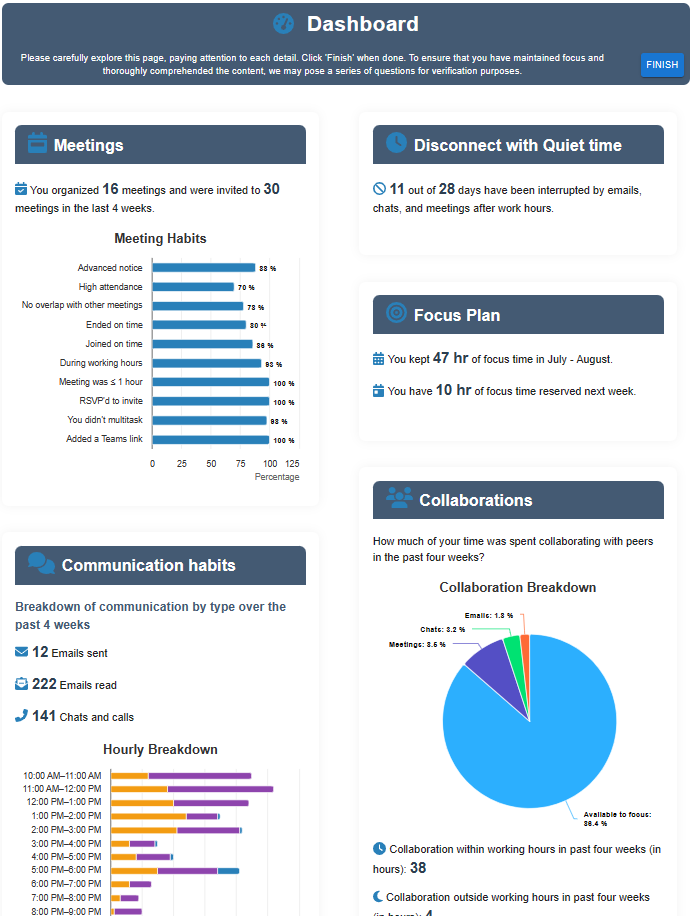}
        \caption{Dashboard visualization}
        \label{fig:dashboard_viz}
    \end{subfigure}
    \begin{subfigure}[b]{0.5\textwidth}        \includegraphics[width=\textwidth]{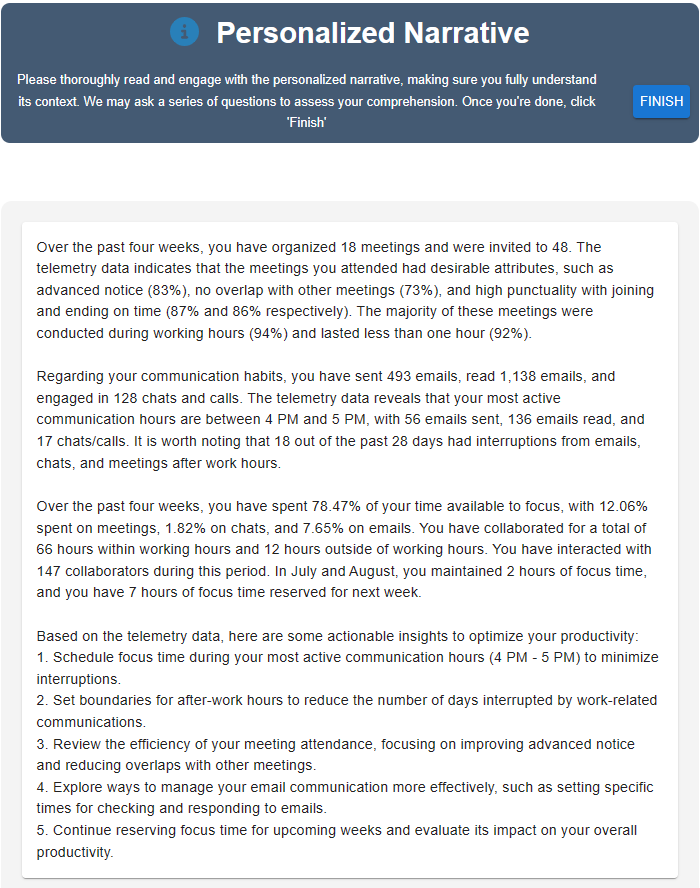}
        \caption{Personalized narrative}
        \label{fig:personalized_narrative}
    \end{subfigure}%
        
    \begin{subfigure}[b]{0.7\textwidth}
        \includegraphics[width=\textwidth]{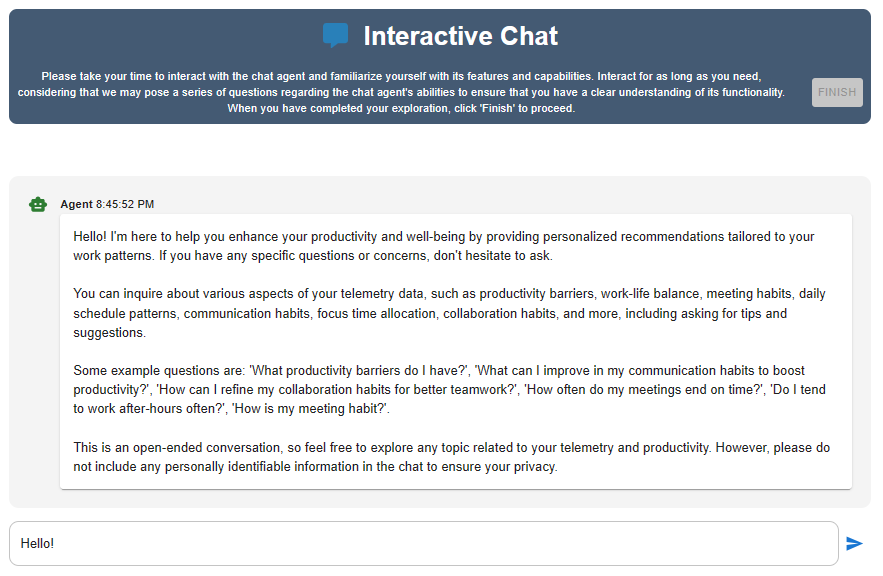}
        \caption{AI productivity agent}
        \label{fig:interactive_agent}
    \end{subfigure}
   
    \caption{Designing Productivity Solutions: Figure (a) displays the dashboard presentation, (b) shows the AI-based personalized narrative, and (c) the AI productivity agent.}
    \label{fig:three_modes_presentation}
\end{figure}

\para{Designing Productivity Solutions:} Having addressed these technical challenges, we proceeded with designing three distinct productivity solutions: an AI productivity agent (also referred to as an interactive chat agent), a personalized narrative (generated via GPT-4), and a traditional visual dashboard. We began by integrating telemetry data directly into the AI agent's system prompt, a crucial step that empowered the agent with the ability to analyze and respond based on individual user behaviors and patterns. This data-driven approach ensured that each participant received feedback and suggestions that reflected their unique work habits. We then drew from the insights gathered during the first phase of our study. The preferences and expectations participants shared during this phase became the foundation for designing the system prompt of the AI agent. Parallel to the development of the AI agent, we embarked on creating an AI-based personalized narrative, an alternative means of communicating productivity insights. This narrative was also crafted to align with the specific preferences participants had voiced. Beyond just these preferences, we introduced a structured format to this narrative, ensuring as consistent delivery of insights as we could across participants. The narrative format presented information in a structured, story-like manner, providing users with an overview of their productivity patterns and suggesting improvements. In contrast, the interactive agent engaged users in a conversational interface, offering personalized guidance based on user inputs and queries. This interactive nature allowed for a more tailored and dynamic exchange, though it was limited to offering advice rather than executing tasks directly. To perfect the prompts for both the AI productivity agent and the personalized narrative, we engaged in multiple cycles of testing and refinement. Each iteration was an opportunity to align more closely with user needs and preferences. Refer to Appendix ~\ref{sec:prompts_agents} for the full prompts used for the productivity agent and personalized narrative. For the development of the dashboard visualization, we recognized that many participants had established familiarity with Viva Insights, so we aimed to make our dashboard's visual presentations echo those of Viva Insights. While GPT-4 was used to extract the telemetry data that populated the dashboard, the dashboard itself was a traditional visualization tool using standard charts and metrics rather than involving any generative AI components. We designed a range of visual representations, from bar charts that provided at-a-glance insights to pie charts that broke down data into comprehensible segments. Furthermore, we chose to present certain telemetry data numerically, ensuring users had access to precise figures. Figure~\ref{fig:three_modes_presentation} presents the AI productivity agent along with the other two presentation modes. 

\para{Evaluation Procedure}: The evaluation process began with participants using our data scraper tool to collect their Viva Insights information, which was then shared with us upon their review. Each presentation mode incorporated this high-level telemetry data from individual participants, allowing for the provision of insights (in the case of the dashboard) and tailored assistance and recommendations (in the case of both the personalized narrative and the AI productivity agent) to improve productivity. Consequently, participants interacted with the three presentation modes, seeking guidance on enhancing their productivity based on their unique work patterns and habits. The study flow is outlined below:

\begin{enumerate}
\item Participants used the scraper tool to gather and share the telemetry data with the research team, identified by a unique code assigned to each participant. The tool submitted three types of files to our storage: a) the raw telemetry data, b) the telemetry data transformed to JSON format using GPT-4, and c) the personalized narrative generated using GPT-4.
\item We provided each participant with a Qualtrics survey link with their unique code embedded in it. This survey included basic demographic questions and links to access the three presentation modes.
\item To mitigate potential ordering effects, participants engaged with the three presentation modes in a randomized sequence and evaluated each mode. The survey displayed the links to the websites featuring the dashboard, interactive chat agent, or personalized narrative in a random order. When participants clicked on the links, the web pages opened with the participant's unique code embedded as part of the query parameter.
\item Since the study aimed to assess the duration of engagement with each mode, participants were allowed unlimited time for exploration. However, each presentation mode instructed participants to familiarize themselves with the presentations' capabilities and insights, as questions about their functionality might be asked. This ensured full engagement with the chosen presentation mode. As the presentation mode had access to the participant's unique code as part of its query parameter, it could refer to the participant's data in our storage. For the dashboard, this meant reading the JSON-transformed telemetry data of the participant and using it to populate the visualizations. For the personalized narrative, it would load the relevant pre-generated narrative, ensuring that participants did not need to wait for their personalized narrative to be generated once they opened the website. For the interactive chat agent, this meant reading the telemetry data and providing it as part of the system prompt to the chat agent. To complete their interaction, participants clicked the "Finish" button available on top of each page, and the duration of their engagement was recorded for further analysis. After clicking "Finish," they were presented with a passcode that they entered into their survey to proceed to the next presentation mode. Access to the next mode was granted only upon submission of the correct passcode. After entering the correct passcode, participants answered questions related to the current presentation mode before advancing to the next one. We maintained consistent page design for each of the three presentation modes, i.e., the buttons looked the same and were placed in the same position, and the colors, spacing, and other visual elements remained the same.
\item After completing all three modes and evaluating them through a series of survey questions, participants ranked the modes according to preference and provided any pros, cons, or suggestions. See Appendix~\ref{sec:user_experience} for the list of questions asked during the evaluation.
\end{enumerate}

\para{Demographics: }Appendix~\ref{sec:demographics_secondphase} presents the demographic information of our participants. Most of the participants identified as men (60.0\%, N=24), while 40.0\% (N=16) identified as women. In terms of age group, the largest proportion of participants was in the 26-35 (30.0\%, N=12) and 46-55 (32.5\%, N=13) age ranges, followed by 36-45 (17.5\%, N=7), 56-65 (12.5\%, N=5), and 18-25 (7.5\%, N=3) age ranges. Regarding job roles, the majority of participants worked in product management (52.5\%, N=21), with software development/engineering being the second most common role (10.0\%, N=4). Other roles included administrative/operations (7.5\%, N=3), data science/analytics (7.5\%, N=3), customer support (5.0\%, N=2), IT/infrastructure (2.5\%, N=1), and other roles (15.0\%, N=6).

%% file: 5_corelation.tex
\section{Results}
\label{sec:results}
In this section, we provide an overview of the results obtained from both phases of our study: the user surveys and the design and evaluation of the productivity agent.
\subsection{Exploring User Preferences for Productivity Agents}
\label{sec:phase1_userpreferences}
In the first phase of the study, we conducted a comprehensive user survey to gather insights into participants' preferences and needs concerning a personalized productivity agent. The survey aimed to explore various aspects of productivity, communication style, agent's approach, personality traits, personalization, and privacy. We divide the results into themes based on the main topics of the survey questions to better understand the participants' perspectives and tailor the productivity agent we design in the second phase to meet their specific requirements effectively. Please refer to Appendix ~\ref{sec:user_survey_responses} for the full list of responses. 

\para{\textit{Productivity Perspectives.}}
We focused on understanding participants' perspectives on productivity, the usefulness of a productivity agent, the agent's personality traits, and the areas in which they would like assistance. The large majority of participants (95.6\%) reported that making progress or completing scheduled tasks would make them feel productive at the end of the day. Solving novel, unexpected tasks (76.6\%) and helping a colleague accomplish a task (72.2\%) were also significant contributors to their feelings of productivity. Participants could select multiple options for this question.

  \begin{figure}[ht!]
    \centering
    \begin{subfigure}[b]{0.5\textwidth}
    \includegraphics[width=\textwidth]{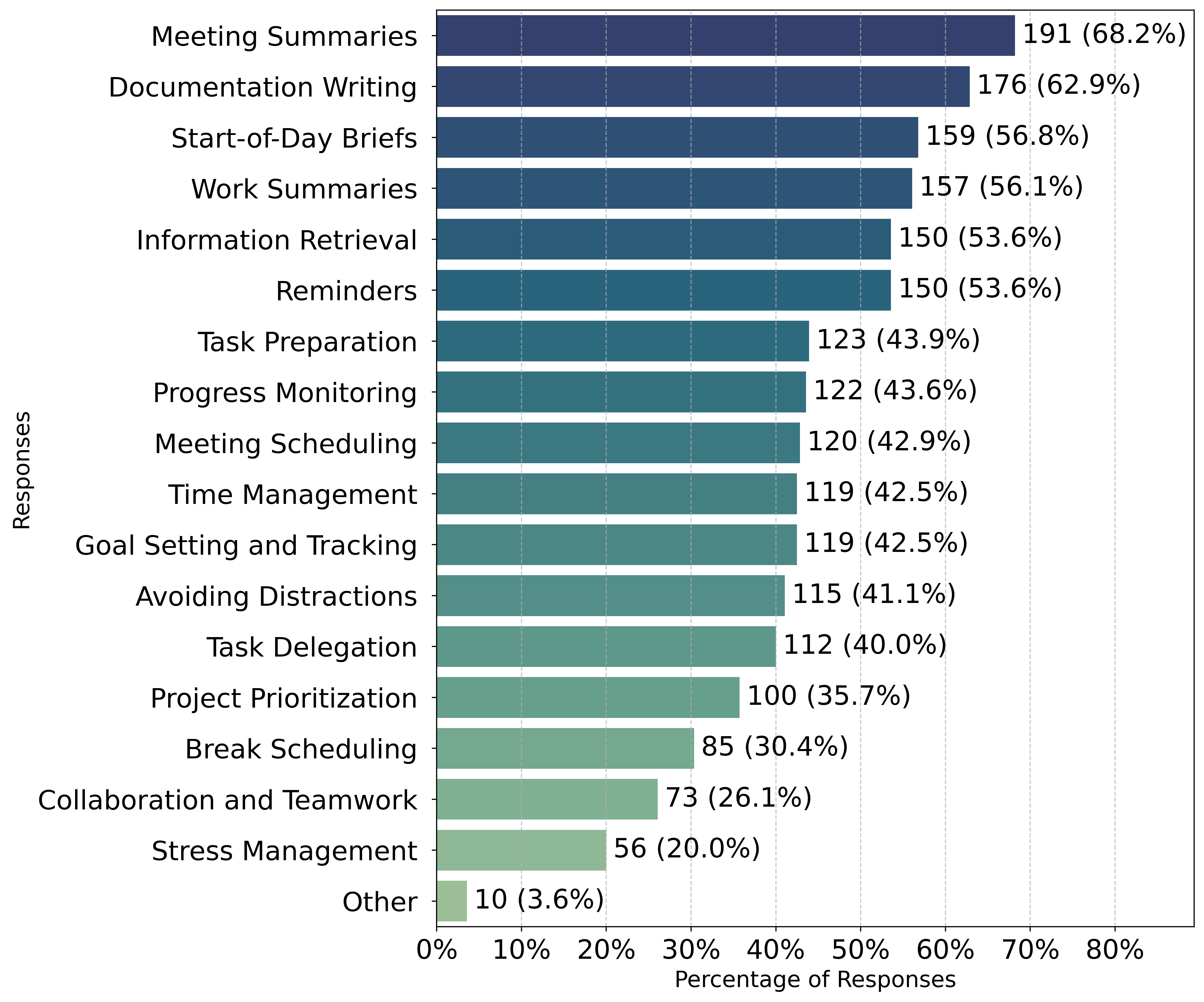}
        \caption{Desired areas of help with productivity agent}
        \label{fig:desired_productivity_areas}
    \end{subfigure}%
    ~
    \begin{subfigure}[b]{0.5\textwidth}
        \includegraphics[width=\textwidth]{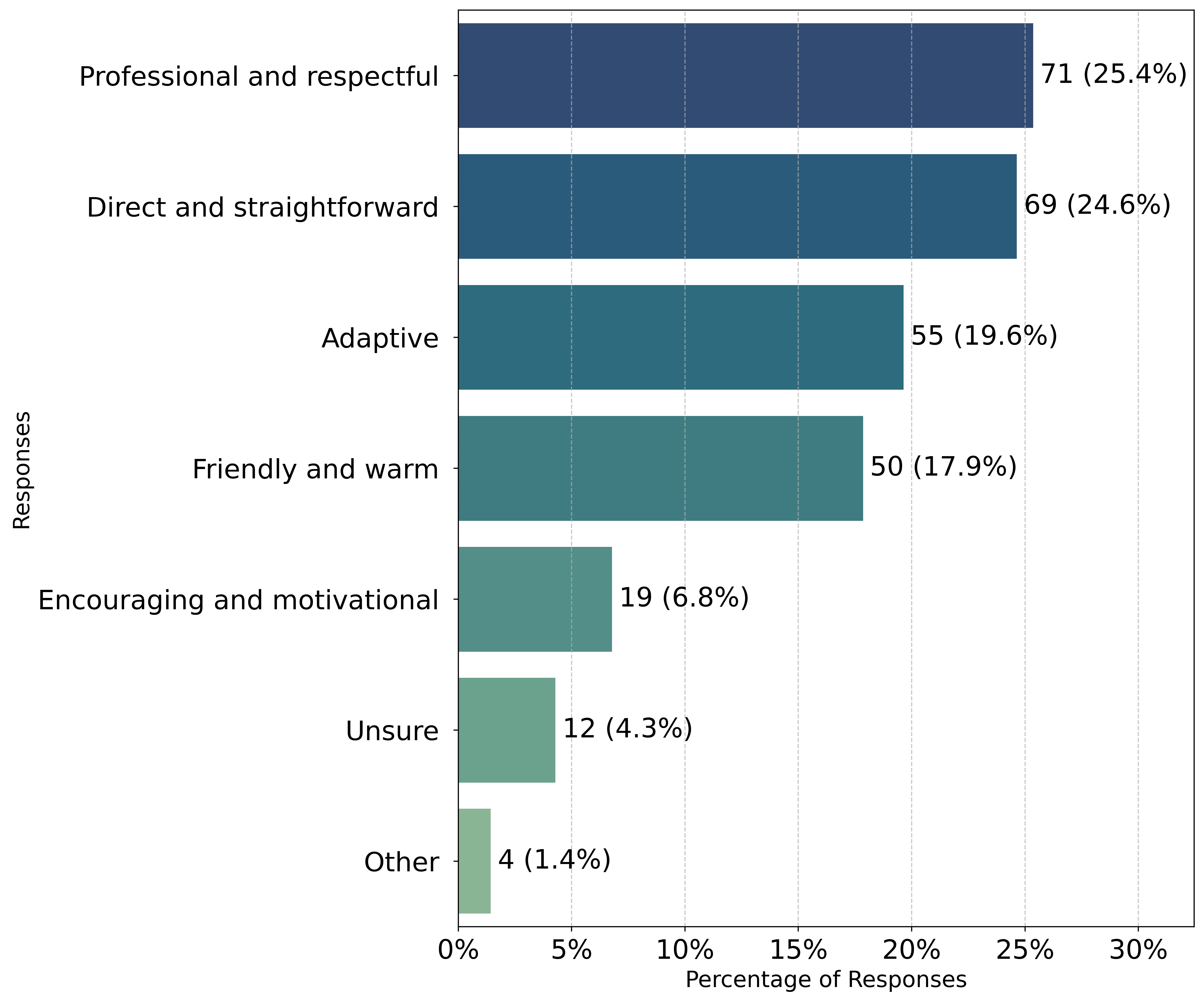}
        \caption{Preferred language or tone}
        \label{fig:preferred_language}
    \end{subfigure}
    \caption{Productivity Areas and Language: Figure (a) illustrates the area that the participants struggle with or would like an AI productivity agent to help them with. Figure (b) shows the preference of participants with regard to the agent's language or tone.}
\end{figure}

Out of all the respondents, 280 (77.1\%) expressed that a productivity agent would be beneficial to them. Consequently, we presented the remaining survey questions only to these participants, who expressed a need for a productivity agent. Among them, 171 (61\%) believed that the personality of the productivity agent would have a moderate-to-significant impact on its effectiveness as a productivity tool. As shown in Figure~\ref{fig:desired_productivity_areas}, when asked about the areas in which participants would like an AI agent to provide assistance, the top five categories were: 191 meeting summaries (68.2\%), 176 documentation writing (62.9\%), 159 start-of-day briefs (56.8\%), 157 work summaries (56.1\%), and 150 information retrieval (53.6\%). These findings provide valuable insights into the specific tasks and areas where a personalized productivity agent can offer the most significant benefits to its users.

Note that in our study's second phase, we consciously selected productivity-enhancing features such as time management, meeting scheduling, and work summaries for our AI agent's focus. This decision was based on substantial user interest and the suitability for immediate application and testing within diverse work settings. These areas, while not the highest in demand according to our Phase 1 findings, presented a unique opportunity to evaluate the real-world efficacy of AI-driven productivity advice. Furthermore, the inclusion of these features enabled a comparative analysis with traditional data presentation methods like dashboards and narrative summaries. This comparison was instrumental in assessing user experience and preference, providing a clearer understanding of the potential and limitations of AI agents in enhancing productivity. The insights gained here are invaluable for directing the evolution of AI productivity tools, ensuring they align more closely with user needs and preferences in subsequent developments.

\para{\textit{Communication Style.}}  We explored participants' preferences regarding the tone and style of communication, as well as their desired level of rapport-building with the productivity agent. A total of 99 participants (35.4\%) expressed their preference for a communication tone that combines elements of both casual and formal styles, maintaining a friendly demeanor while still conveying professionalism. Additionally, 95 participants (33.9\%) mentioned that their preferences for communication tone varied depending on the specific task or situation. As shown in Figure~\ref{fig:preferred_language}, when it came to rapport-building, 71 participants (25.4\%) responded that a professional and respectful language or tone would make them feel most comfortable opening up to the productivity agent about concerns related to productivity. Similarly, 69~participants (24.6\%) preferred a direct and straightforward tone.

Regarding the style of messages, 99 participants (35.4\%) responded that they would prefer direct and to-the-point messages from the productivity agent, while 86 (30.7\%) said that they would prefer a mixture of chatty and conversational as well as direct and to-the-point. These findings highlight the importance of incorporating a balance between casual and formal communication styles, as well as the need for adaptability in tone, depending on the situation and user preferences.

\begin{figure}[ht!]
    \centering
    \begin{subfigure}[b]{0.5\textwidth}        \includegraphics[width=\textwidth]{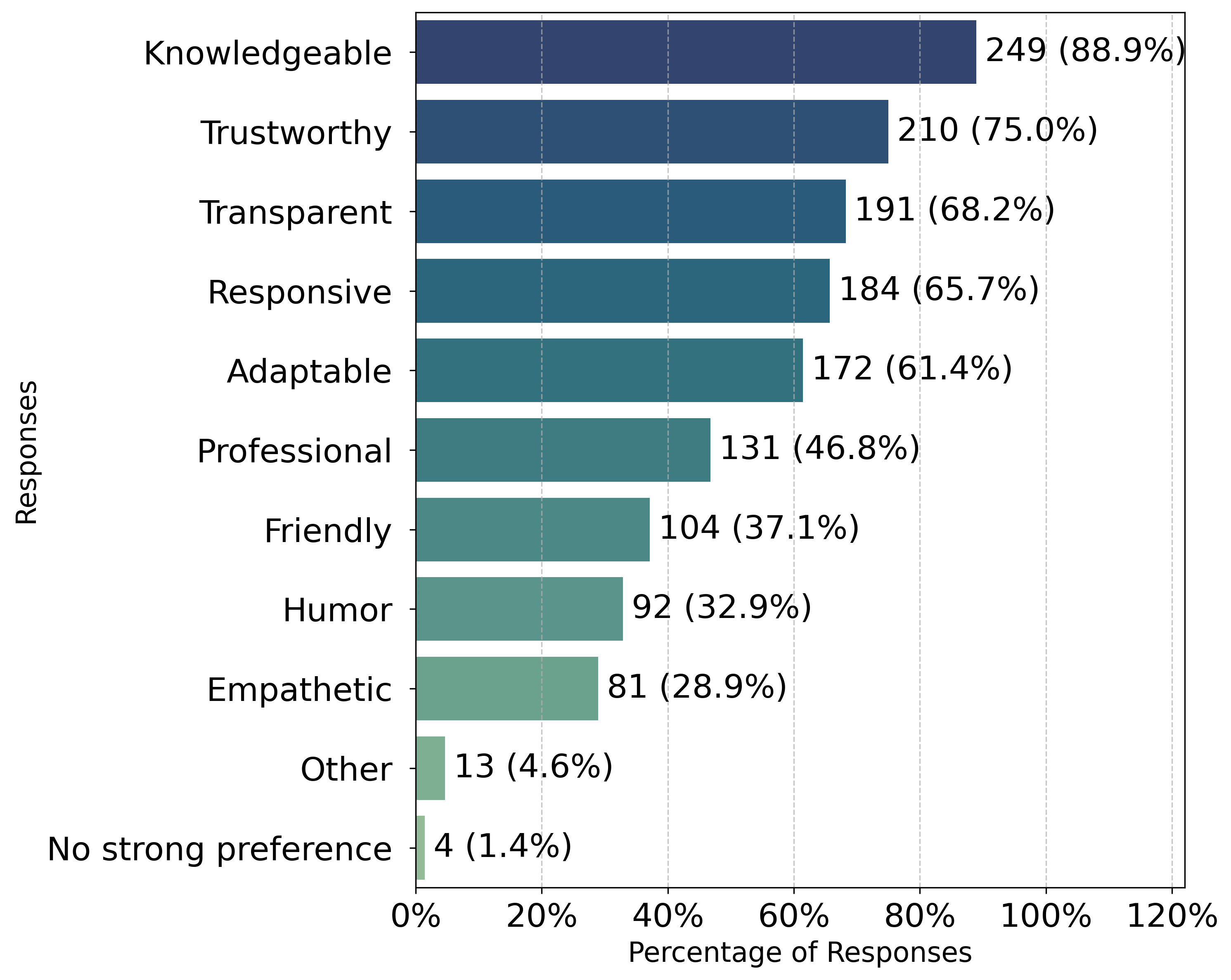}
        \caption{Desired personality traits}
        \label{fig:personality_traits}
    \end{subfigure}%
    ~
    \begin{subfigure}[b]{0.5\textwidth}
        \includegraphics[width=\textwidth]{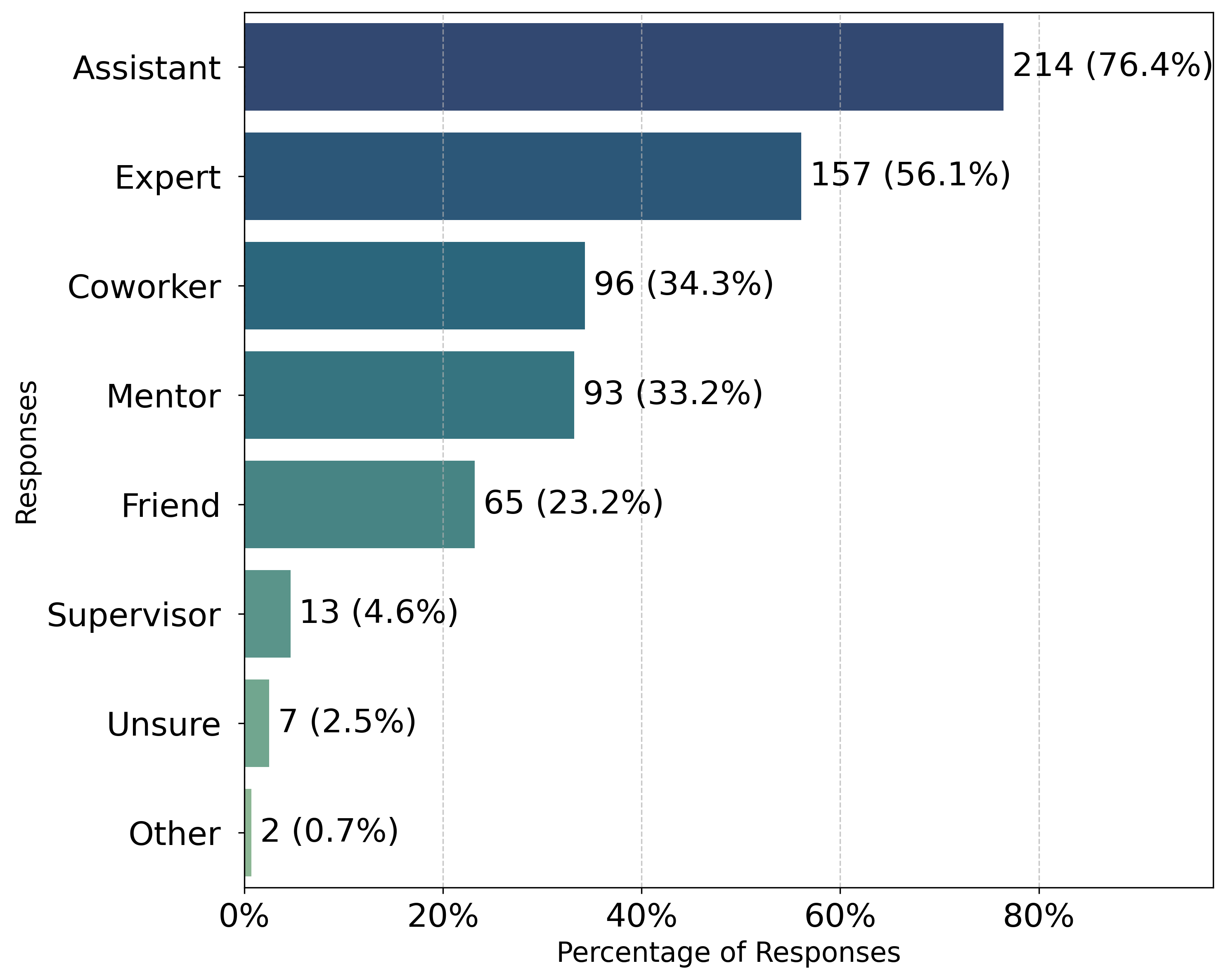}
        \caption{Preferred social roles}
        \label{fig:social_roles}
    \end{subfigure}
    \caption{Personality Traits and Social Roles: Figure (a) depicts the different personality traits participants prefer in the productivity agent. Figure (b) shows the social role desired by the participants for the agent.}
\end{figure}

\para{\textit{Approach \& Personality Traits.}}
We investigated the participants' preferences regarding the productivity agent's approach to task assistance, desired personality traits, and preferred social roles.

Most of the participants, 169 (60.4\%), expressed their preference for the agent to provide proactive suggestions to enhance productivity. In contrast, 77 participants (27.5\%) wanted the agent to provide assistance and suggestions only when explicitly asked for help. The agent's preferred personality traits are depicted in Figure~\ref{fig:personality_traits}; the most common preferences were: knowledgeable (88.9\%), trustworthy (75.0\%), and transparent (68.2\%). Similarly, Figure~\ref{fig:social_roles} represents the social role that people desired for the agent. The top choices were that of an assistant (76.4\%) and expert (56.1\%). The supervisor role was the least preferred, with only 4.6\% of participants opting for it. 

\para{\textit{Personalization \& Privacy.}}
We explored participants' preferences regarding personalized recommendations and privacy concerns when using a productivity agent.

A significant majority of participants, 209 (74.6\%), expressed that they would prefer a moderate to high level of personalization in the productivity agent. This enables the agent to provide highly personalized advice tailored to their specific needs, considering various aspects of their work habits and preferences. In terms of privacy, 152~participants (54.3\%) felt either somewhat comfortable or completely comfortable sharing their personal information with the productivity agent. However, 58 participants (20.7\%) expressed discomfort in sharing their personal information. Interestingly, a majority of participants, 205 (73.2\%), were either somewhat comfortable or completely comfortable sharing high-level telemetry data for enhancing the quality and guidance provided by the productivity agent.

\para{\textit{Negative traits to avoid.}}
We asked participants an open-ended question -- \textit{``In your opinion, what are some potential negative personality traits that should be avoided in a productivity agent to ensure a positive, effective, and supportive user experience?''} The responses emphasized the importance of avoiding traits such as nagging, aggressiveness, pushiness, verbosity, overly casual or friendly demeanor, judgmental or patronizing tones, and condescension. Participants also warned against providing false or exaggerated praise, taking on a ``taskmaster'' or micromanager role, and presenting inaccurate information. Other concerns included anthropomorphism, avoiding complex language or exclusionary phrasing, and being too intrusive. Participants generally preferred an agent that acts as a helpful assistant, rather than a replacement for a counselor or a parent.
\begin{table*}[ht!]
\caption{Design of the AI Productivity Agent Based on Phase 1 Survey Findings.}
\smaller
\begin{tabular}{@{}p{0.25\linewidth}p{0.35\linewidth}p{0.35\linewidth}@{}}
\textbf{Theme} & \textbf{User Preference (Phase 1)} & \textbf{Integration in Productivity Agent (Phase 2)} \\ \midrule

\rowcolor[HTML]{EFEFEF}
Productivity Perspectives & Participants felt productive when making progress on scheduled tasks, solving unexpected tasks, and helping colleagues & Agent designed to offer strategies for task management, prioritizing, and collaborative assistance. It is developed to help users achieve an optimal work-life balance and improve their productivity. \\

Communication Style & Preference for a tone that balances casual and formal styles; adaptability in tone based on task and situation; direct and straightforward communication rather than chatty & Agent uses a professional yet friendly tone, adaptable to the context and user's current task. It is designed to communicate concisely without being overly chatty or verbose. \\

\rowcolor[HTML]{EFEFEF}
Approach & Desire for proactive suggestions and assistance to enhance productivity & Agent designed to provide proactive productivity suggestions without being too intrusive and pushy \\

Personality Traits & Valued traits included being knowledgeable, trustworthy, and transparent. & Agent developed to demonstrate these traits consistently across interactions.  \\

\rowcolor[HTML]{EFEFEF}
Personalization & Preference for a moderate to high level of personalization, balancing with privacy concern & Agent uses high-level telemetry data to personalize suggestions while ensuring user privacy \\

Negative Traits to Avoid & Avoidance of traits like being nagging, aggressive, pushy, overly casual, judgmental, or patronizing. & Agent prompted to avoid these negative traits, focusing on supportive and informative interactions. \\

\bottomrule
\end{tabular}
\label{tbl:ai_agent_design_based_on_user_preferences}
\end{table*}

\subsection{Design of the AI Productivity Agent Informed by User Preferences and Telemetry Data}

The development of our GPT-4-powered productivity agent focused on two key aspects: effective integration of telemetry data for personalization and alignment with user preferences identified in Phase 1 (Section \ref{sec:phase1_userpreferences}). The agent's system prompt was dynamically constructed to include the user's telemetry data, structuring it into clear sections about meeting habits, communication patterns, focus time, and work-life balance metrics. We formatted the telemetry data to enable the agent to make specific comparisons and identify patterns, such as \textit{``Your meetings are 45\% longer than the average''} or \textit{``You have 11 days with after-hours work in the past month.''} The agent was designed to reference this data explicitly in its responses, grounding its recommendations in the user's actual work patterns.

Based on Phase 1 findings, we configured the agent to maintain a professional yet friendly tone (preferred by 35.4\% of survey participants), provide proactive suggestions while respecting user autonomy (aligned with 60.4\% of participants' preferences), and demonstrate traits of being knowledgeable and trustworthy (preferred by 88.9\% and 75.0\% of participants respectively). A summary of these findings and their corresponding implementations in the AI agent's design is presented in Table \ref{tbl:ai_agent_design_based_on_user_preferences}. The table illustrates how each identified theme from the survey was translated into specific features of the AI productivity agent. The agent was specifically instructed to avoid characteristics identified as negative in our survey, such as being pushy or overly casual (See agent's system prompt in Appendix \ref{sec:prompts_agents}).

Additionally, personalization emerged as a crucial aspect, with users expressing a need for tailored recommendations balanced against privacy concerns. Accordingly, the agent leverages high-level telemetry data to offer personalized suggestions, carefully designed to respect user privacy. When a user asks about improving their communication habits to boost productivity, the agent analyzes their personal communication metrics from the telemetry data, identifies specific patterns, and provides recommendations grounded in their actual behavior. For example, as shown in Appendix \ref{sec:sample_resposne_aiagent} Figure \ref{fig:sample_chat3}, when a user asks, \textit{``What can I improve in my communication habits to boost productivity?''}, the system provides personalized recommendations based on their specific work patterns. The agent identifies that the user engages heavily in email and communication during their peak productivity hours (12-3 PM) and suggests redistributing these activities to less productive times. The system also recognizes work-life balance issues by noting after-hours communication occurring in 11 out of 28 days. This example demonstrates how telemetry data enables meaningful personalization through analysis of broader work patterns—identifying meeting scheduling trends, communication habits, focus time preferences, and work-life balance patterns—to provide targeted recommendations for productivity improvement.

\subsection{Productivity Solution Showdown: AI Agent vs. Dashboard vs. Personalized Narrative}
In this section, we present the findings from the second phase of our study. Specifically, we explore the interaction duration, participants' overall experiences, anticipated usage frequency, and preference rankings for the three presentation methods: the AI productivity agent, visual dashboard, and AI-based personalized narrative. 
\begin{figure}[ht!]
\centering
   \begin{subfigure}{0.5\linewidth} \centering
\includegraphics[width=\linewidth]{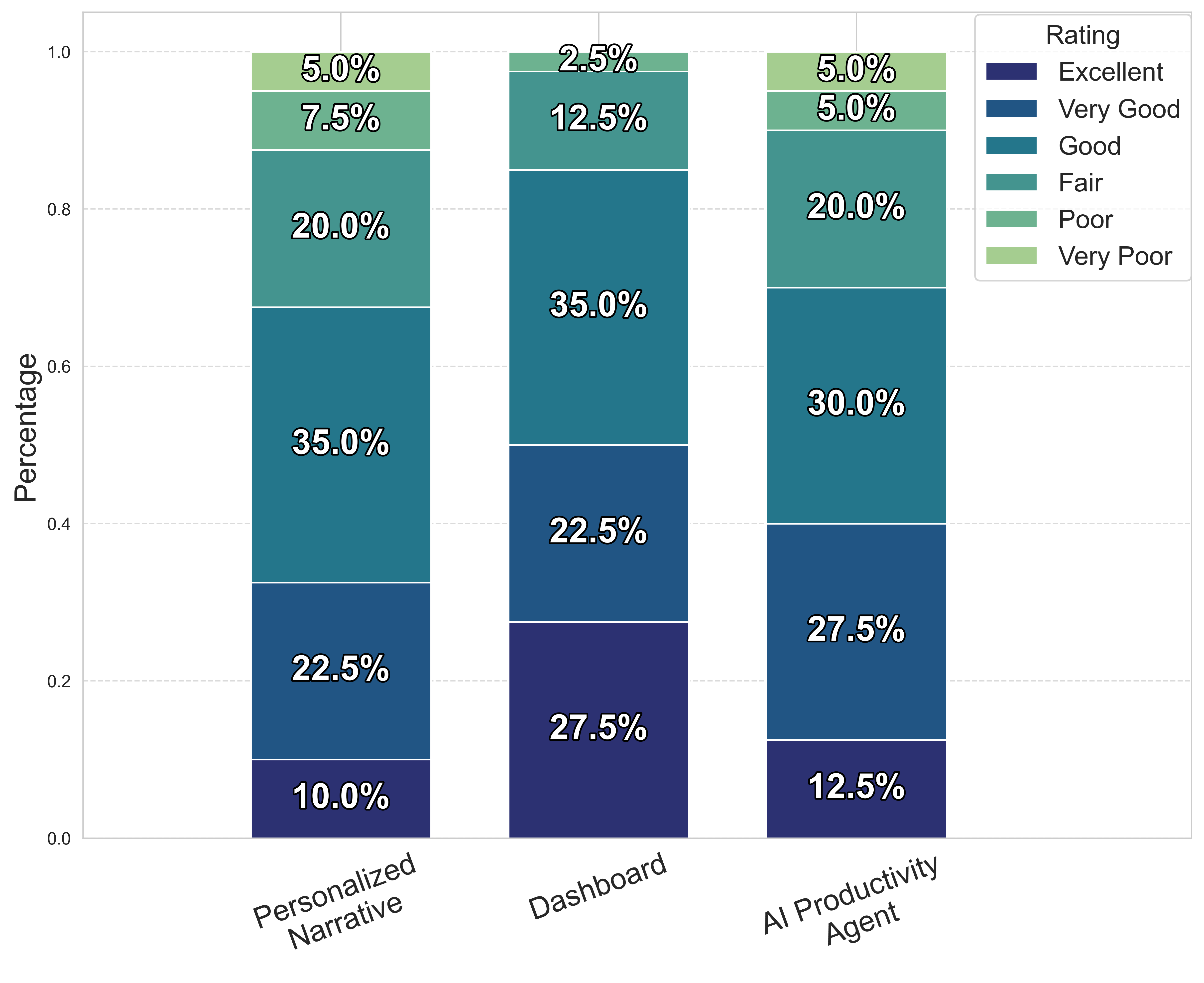}
     \caption{Ratings of overall experience}\label{fig:overallexpa}
   \end{subfigure}
   ~
   \begin{subfigure}{0.5\linewidth} \centering
\includegraphics[width=\linewidth]{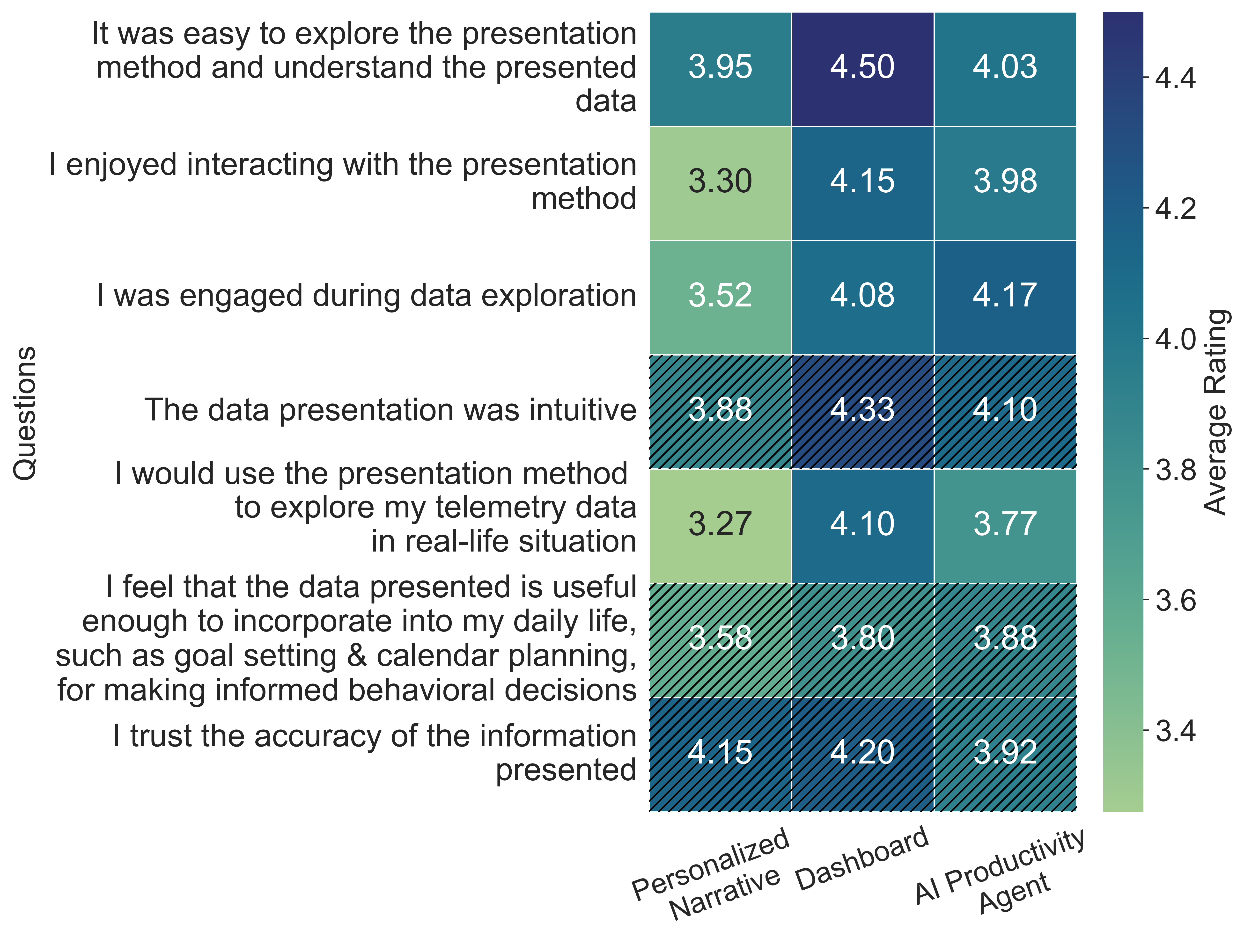}
     \caption{Average rating of presentation methods across different criteria.}\label{fig:overallexpb}
   \end{subfigure}
\caption{Overall Experience: Figure (a) illustrates the participants' overall experience ratings for three data presentation methods: Personalized Narrative, AI Productivity Agent, and Dashboard. Figure (b) compares the ratings of these methods across various criteria. We performed the Kruskal Wallis test to analyze the differences between the three groups in Figure (b). Criteria without any statistically significant differences (i.e., p-value > 0.10) are marked with a crosshatch shading pattern - a series of intersecting lines drawn at an angle to create texture.} 
\label{fig:overall_experience}
\end{figure}

\begin{figure}[ht!]
\centering
   \begin{subfigure}{0.5\linewidth} 
\includegraphics[width=\linewidth]{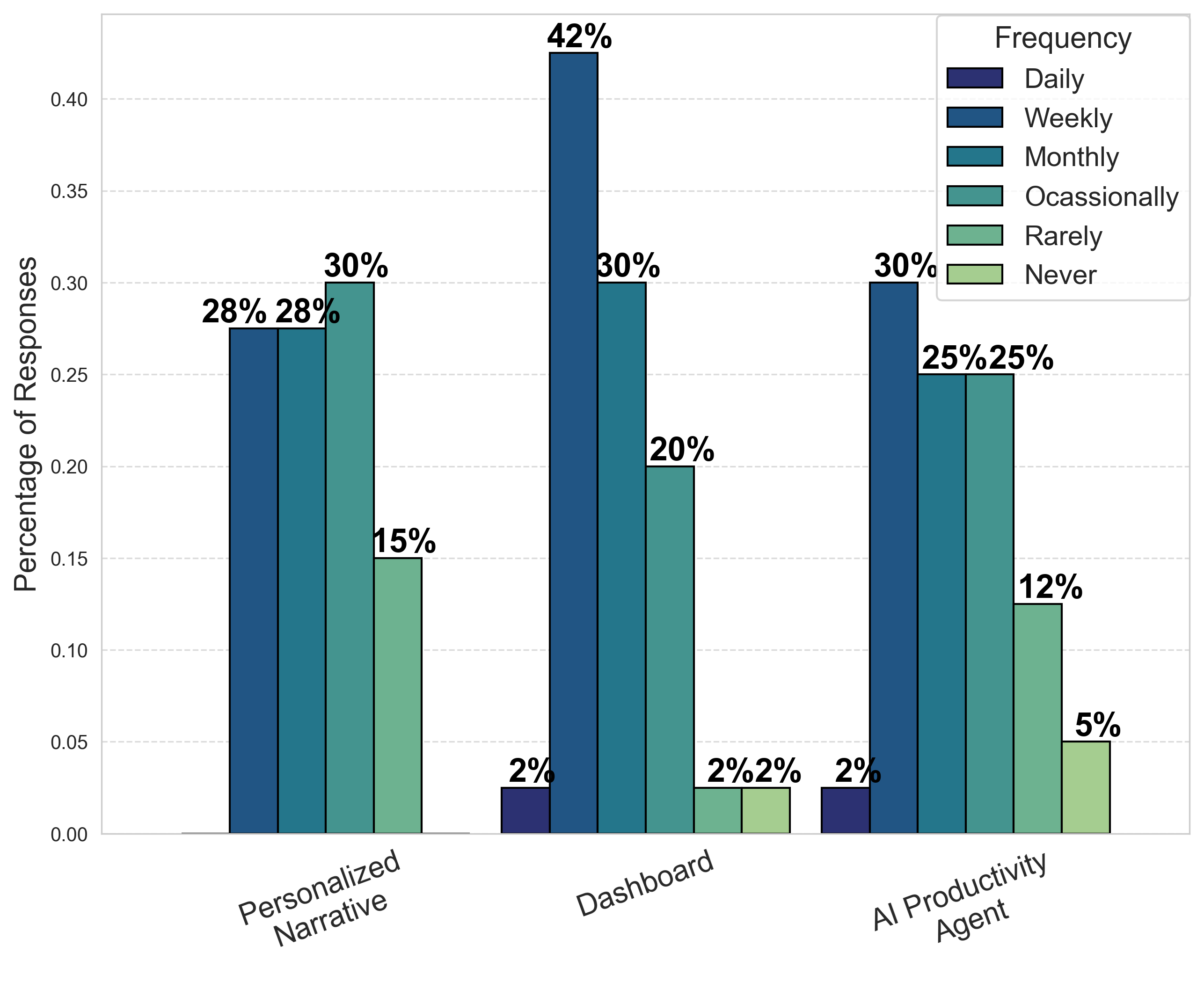}
\caption{Anticipated usage frequency}\label{fig:anticipateduse}
   \end{subfigure}
   ~
   \begin{subfigure}{0.5\linewidth} 
\includegraphics[width=\linewidth]{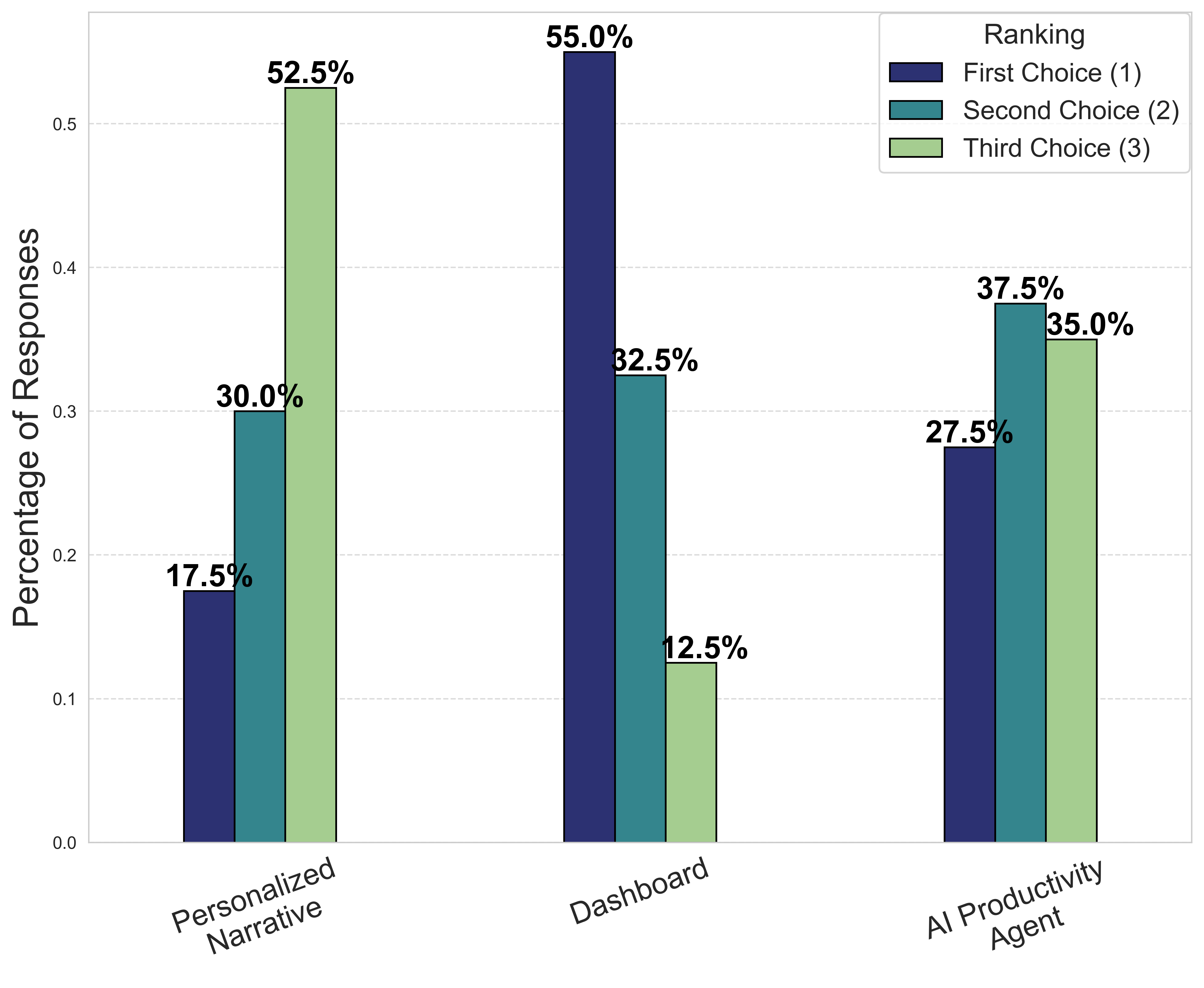}
     \caption{Ranking based on preferences}\label{fig:ranking}
   \end{subfigure}
\caption{Anticipated use and Ranking: Figure (a) displays the anticipated usage of the three distinct data presentation methods. In contrast, Figure (b) showcases the participants' rankings of these methods based on their preferences, with a ranking of 1 indicating the most preferred method and 3 signifying the least preferred method.} 
\label{fig:use_and_preference}
\end{figure}

Participants spent an average of 60.58 seconds (with a standard deviation of 38.98 seconds) interacting with the dashboard. In comparison, they spent considerably more time on the AI productivity agent, with an average duration of 315.48 seconds (standard deviation: 272.23 seconds). The personalized narrative had a similar interaction time to the dashboard, with an average of 65.21 seconds (standard deviation: 55.39 seconds). After interacting with each method, participants were asked to rate their overall experience on a six-point Likert scale ranging from Very Poor to Excellent. As shown in the stacked bar graph in Figure~\ref{fig:overallexpa}, 26.2\% of participants rated the Dashboard as excellent, while only 11.9\% and 9.5\% rated Excellent for the productivity agent and personalized narrative, respectively. Combining the top three experiences (excellent, very good, and good), 81\% considered the Dashboard good or above, 66.7\% considered the AI productivity agent good or above, and 64.2\% considered the personalized narrative to be good or above. To compare user ratings across the three methods, we conducted a chi-square test of independence. Given our relatively small sample size, we used a significance level of $\alpha$ = 0.10 to account for reduced statistical power. The analysis revealed statistically significant differences in the distribution of `Excellent' ratings across conditions ($\chi$\textsuperscript{2} = 5.16, p-value = 0.07), suggesting that participants had different preferences when giving `Excellent' ratings to the three conditions. We did not find any significant differences in the distribution of other rating categories (Very Good, Good, Fair, Poor, Very Poor) across the methods. Following the broader question, participants were asked to rate their experience on specific criteria for each method using a five-point Likert scale from Strongly Disagree (1) to Strongly Agree (5). In Figure~\ref{fig:overallexpb}, a heatmap displays the average scores obtained by each method for these questions. Most darker blocks (in blue) are centered on the Dashboard, indicating higher average scores for the ease of exploring and understanding data, the use of the presentation method in real-life situations, and enjoyment in interaction. The boxes with crosshatch shading patterns (intersecting lines) indicate that the differences were not statistically significant (Kruskal Wallis test; p-value > 0.10) among the three methods for that particular criterion. The productivity agent scored better in only one area: engagement during data exploration. The personalized narrative performed the worst with the lowest scores among the three. Regarding anticipated usage frequency, we show in Figure~\ref{fig:anticipateduse} that 42\% of participants selected weekly for the dashboard and 30\% selected weekly for the productivity agent, which was the majority score for both groups. For personalized narrative, 28\% of participants said that they would use it weekly and the same amount also said they would rather use it monthly.  30\% of the participants said they would use the personalized narrative occasionally. However, we did not obtain any statistically significant differences in the distribution of usage frequencies across the three methods. Lastly, participants ranked the three methods in order of preference (1 being the most preferred and their first choice, 3 being the least preferred and their last choice). As shown in Figure~\ref{fig:ranking}, 55\% rated the Dashboard as their most preferred method, followed by 27.5\% preferring the productivity agent and 17.5\% preferring the personalized narrative. Over half of the participants (52.5\%) rated the personalized narrative as their least preferred method. A chi-square test of independence revealed significant differences in order of preference across the three methods for rank 1 i.e. most preferred method ($\chi$\textsuperscript{2} = 13.57, p-value < 0.01). The least preferred method i.e. rank 3 also showed significant differences ($\chi$\textsuperscript{2} = 14.47, p-value < 0.01), but not rank 2.

\subsection{AI Productivity Agent Performance}
In this section, we look into the performance and capabilities of the AI productivity agent, examining the aspects of its interaction with users. We analyze the perceived capabilities of the agent, chat metrics, and productivity-related themes that emerged from users' questions. Our goal is to gain insights into the effectiveness of the AI productivity agent.

\para{\textit{Agent Capabilities.}}
We designed the productivity agent based on the findings from our user survey and aimed to align it effectively with the insights from the first phase. To assess this alignment, we asked participants specific questions regarding the interactive chat agent's perceived abilities. It is important to note that these questions were posed after participants had indicated their preferred mode of presentation, ensuring that their rankings were not influenced by these queries. For a comprehensive overview of the capabilities and participants' ratings of the agent's performance, please refer to Table~\ref{tbl:agent_capabilities}.

\begin{table*}[ht!]  
\caption{Perceived capabilities of the interactive agent. The responses are categorized as follows: SD for Strongly Disagree, SWD for Somewhat Disagree, NAND for Neither Agree Nor Disagree, SWA for Somewhat Agree, SA for Strongly Agree, and N/A for Non-applicable. We bold the group that has the majority of the responses -- either Agree (SA, SWA) or Disagree (SD, SWD) or Neutral (NAND, N/A).}  
\small
\begin{tabular}{@{}p{0.4\linewidth}cccccc@{}}  
\textbf{Capabilities} & \multicolumn{6}{c}{\textbf{Count}} \\  
                             & \textbf{SD} & \textbf{SWD} & \textbf{NAND} & \textbf{SWA} & \textbf{SA} & \textbf{N/A} \\ \midrule   
\rowcolor[HTML]{EFEFEF}The agent provided useful tips and strategies for improving productivity & 1 (2.5\%) & 1 (2.5\%) & 6 (15\%) & \textbf{15 (37.5\%)} & \textbf{17 (42.5\%)} & 0 (0\%) \\  
The agent was able to understand my productivity challenges &  0 (0\%) & 1 (2.5\%) & 8 (20\%) & \textbf{19 (47.5\%)} & \textbf{9 (22.5\%)} & 3 (7.5\%) \\  
\rowcolor[HTML]{EFEFEF}The agent's empathy and tone helped me feel more comfortable discussing my productivity concerns  & 2 (5\%) & 5 (12.5\%) & 10 (25\%) & \textbf{9 (22.5\%)} & \textbf{12 (30\%)} & 2 (5\%) \\  
The agent personalized its suggestions based on my situation and needs & 1 (2.5\%) & 2 (5\%) & 6 (15\%) & \textbf{16 (40\%)} & \textbf{14 (35\%)} & 1 (2.5\%) \\ 
\rowcolor[HTML]{EFEFEF}The agent provided practical solutions to overcome productivity obstacles  & 3 (7.5\%) & 1 (2.5\%) & 5 (12.5\%) & \textbf{16 (40\%)} & \textbf{13 (32.5\%)} & 2 (5\%) \\ 
The agent used a professional and direct language and tone  & 1 (2.5\%) & 0 (0\%) & 0 (0\%) & \textbf{13 (32.5\%)} & \textbf{26 (65\%)} & 0 (0\%) \\ 
\rowcolor[HTML]{EFEFEF}The agent proactively suggested ways to improve my productivity & 2 (5\%) & 0 (0\%) & 5 (12.5\%) & \textbf{17 (42.5\%)} & \textbf{15 (37.5\%)} & 1 (2.5\%) \\ 
The agent was annoying, aggressive or pushy  & \textbf{26 (65\%)} & \textbf{7 (17.5\%)} & 3 (7.5\%) & 1 (2.5\%) & 2 (5\%) & 1 (2.5\%) \\ 
\rowcolor[HTML]{EFEFEF}The agent was being too intrusive  & \textbf{22 (55\%)} & \textbf{11 (27.5\%)} & 4 (10\%) & 2 (5\%) & 0 (0\%) & 1 (2.5\%) \\ 
The agent was overly critical  & \textbf{20 (50\%)} & \textbf{11 (27.5\%)} & 5 (12.5\%) & 1 (2.5\%) & 0 (0\%) & 3 (7.5\%) \\ 
\rowcolor[HTML]{EFEFEF}The agent seems to be able to easily work out what I might want to talk about  & 1 (2.5\%) & 1 (2.5\%) & 10 (25\%) & \textbf{17 (42.5\%)} & \textbf{10 (25\%)} & 1 (2.5\%) \\ 
The agent seems to have a difficult time seeing things from my point of view & \textbf{8 (20\%)} & \textbf{11 (27.5\%)} & 12 (30\%) & 4 (10\%) & 3 (7.5\%) & 2 (5\%) \\ 
\rowcolor[HTML]{EFEFEF}The agent seems to vary its conversational style to accommodate my mood or disposition  & 1 (2.5\%) & 3 (7.5\%) & \textbf{17 (42.5\%)} & 7 (17.5\%) & 1 (2.5\%) & \textbf{11 (27.5\%)} \\ 
The agent seems to follow what I say and accurately reflects its understanding to me & 1 (2.5\%) & 2 (5\%) & 3 (7.5\%) & \textbf{17 (42.5\%)} & \textbf{17 (42.5\%)} & 0 (0\%) \\ 
\rowcolor[HTML]{EFEFEF}The agent demonstrated the traits of being knowledgeable, trustworthy, transparent, and responsive & 1 (2.5\%) & 3 (7.5\%) & 5 (12.5\%) & \textbf{17 (42.5\%)} & \textbf{14 (35\%)} & 0 (0\%) \\ 
The agent's use of telemetry data enhanced its ability to provide personalized and empathetic support & 1 (2.5\%) & 5 (12.5\%) & 3 (7.5\%) & \textbf{17 (42.5\%)} & \textbf{12 (30\%)} & 2 (5\%) \\ \bottomrule  
\end{tabular}  
\label{tbl:agent_capabilities}  

\end{table*}

29 participants (72.5\%) agreed that the use of telemetry data in the agent enhanced its ability to provide personalized and empathetic support. To design the agent as knowledgeable, trustworthy, and transparent based on the user survey, 31 participants (77.5\%) felt that the agent demonstrated these traits. Additionally, 34 participants (84\%) agreed that the agent followed directions well. However, only 8 participants (20\%) found the agent adaptive in its conversation style to accommodate participants' mood, with 17 (42.5\%) remaining neutral and 11 (27.5\%) considering it non-applicable. We performed a Wilcoxon Signed-Rank Test ($\alpha$ = 0.10) to determine if participants significantly agreed or disagreed compared to a neutral response. We found significant differences from neutral responses in all questions except this one i.e., change in conversation style to accommodate participants' mood or disposition (p-value > 0.10). The agent performed well in avoiding negative traits identified in the user survey (we instructed the agent through its system prompt to avoid such negative traits), as 31 participants (77.5\%) disagreed that it was overly critical, 33 (82.5\%) disagreed that it was intrusive, and 33 (82.5\%) disagreed that it was annoying, aggressive, or pushy. Nevertheless, the agent's ability to see things from the user's point of view received mixed responses, with only 19 participants (47.5\%) disagreeing, 7 (17.5\%) agreeing, and 12 (30\%) remaining neutral. Other findings indicated that the agent proactively suggested productivity improvements (80\%), provided practical solutions to overcome productivity barriers (72.5\%), and offered useful tips and strategies (82.5\%). Participants also felt that the agent personalized its suggestions based on their situation and needs~(75\%) and understood their productivity challenges (70\%).

\para{\textit{Productivity Conversations.}}
When examining the AI productivity agent interactions, we found that users exchanged an average of 10 messages. The average response time was 33 seconds.  
Users wrote an average of 9 words per message, while the chat agent provided more detailed responses with an average of 208 words per message. We examined the chat excerpts to better understand the aspects of the agent that resonated with users, as well as areas where improvements could be made to enhance the overall user experience. The majority of the messages were focused on productivity. In particular, we identified five main productivity-related themes from the users' questions:

\begin{figure}[ht!]
     \begin{subfigure}[b]{0.7\textwidth}
\includegraphics[width=\textwidth]{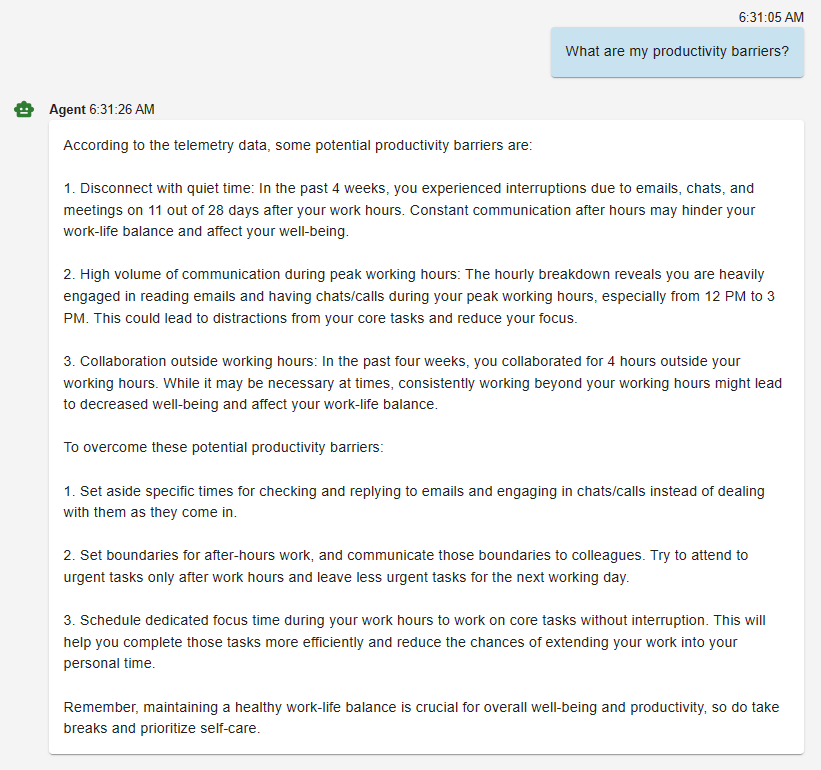}
    \end{subfigure}
    \caption{Sample Conversation: A sample response from the chat agent for the question, \textit{``What are my productivity barriers?''}}
    \label{fig:sample_chat1}
\end{figure}
\begin{enumerate}
    \item Time Management and Work-Life Balance:
Users were concerned about managing their working hours, balancing work across different time zones, and understanding how their work habits compared to their peers. Questions in this theme included: \textit{``How is my work-life balance compared to peers?''}, \textit{``How do you recommend balancing a team based in Pacific Time and Eastern Time with my Central hours?''}, and \textit{``Do you notice a difference in work habits across the 4 weeks?''}
\item Meeting Efficiency and Management:
Participants were interested in improving the efficiency and management of their meetings. They asked about meeting duration, attendance rates, and how to prevent meetings from running over time. Example questions include: \textit{``Why are meetings shorter than 1 hour better for productivity?''}, \textit{``How often do my meetings end on time?''}, and \textit{``What recommendations do you have for ensuring meetings end on time?''}
\item Communication and Collaboration Habits:
Users sought to enhance their communication and collaboration habits to boost productivity. They inquired about strategies to reduce email volume, refine collaboration habits, and improve communication with teammates. Sample questions are: \textit{``What can I improve in my communication habits to boost productivity?''}, \textit{``How can I refine my collaboration habits for better teamwork?''}, and \textit{``How can I be more interactive with my teammates?''}. Figure~\ref{fig:sample_chat1} shows a sample response from the agent for one of these questions.
\item Personalized Insights and Recommendations:
Users aimed to receive tailored insights and recommendations based on their data to improve their productivity. They were interested in understanding their productivity barriers and receiving specific advice. Questions in this theme include: \textit{``What productivity barriers do I have?''}, \textit{``What can I improve in my communication habits to boost productivity?''}, and \textit{``What is one thing I should start doing to be more effective?''}
\item Tools, Templates, and Data Visualization:
Participants expressed interest in tools, templates, and data visualization to streamline their work processes and better understand their productivity. They asked about email templates, note-taking templates, and ways to visualize their data. Example questions include: \textit{``Can you create email templates for common scenario email needs?''}, \textit{``Please share a note-taking template and a template of a concise meeting summary with clear actions?''}, and \textit{``Can you visualize this data for me?''}. Note that the productivity agent we developed did not have the ability to create visualizations.
\end{enumerate}

\section{Strengths, Weaknesses, and Improvement Opportunities}
To gain a deeper understanding of the pros and cons of each method and gather improvement suggestions, we asked participants to provide open-ended feedback. By considering participants' insights and experiences, we can better understand the strengths and weaknesses of the AI productivity agent, visual dashboard, and AI-based personalized narrative, as well as explore opportunities for enhancing their overall performance and user experience.

Based on their feedback, we can say that participants appreciated the interactive chat agent for its ability to provide specific, tailored insights and recommendations. One participant mentioned, \textit{``The ChatBot was amazing!''}, while another said, \textit{``I liked how it suggested solutions for my productivity without me even asking for it.''} However, some participants found it labor-intensive and preferred using it as a secondary step after viewing a dashboard. A participant expressed this sentiment, stating, \textit{``I think the chatbot is a nice tool, but I would prefer to see the data in a visual form first, then maybe use the chatbot for follow-up questions.''} Participants enjoyed the dashboard presentation method for its visual appeal and ease of understanding. They expressed a desire for more interactivity and benchmarking functionality to compare their data against company or organizational averages. A participant commented, \textit{``I really loved the data visualization here, it's exactly what was missing from the chatbot experience. I liked the colors too, they grabbed my attention.''} However, another participant expressed a negative experience, saying, \textit{``While the dashboard looks nice, I feel overwhelmed with all the numbers and charts. I wish there was a simpler way to understand the data.''} Similarly, another participant said, \textit{``The dashboard doesn't provide adequate context for the data, leaving me unsure of how to apply the insights to my situation.''} The personalized narrative method received mixed reviews from participants. Some found it succinct and helpful, while others thought it was too wordy and lacked visual appeal. One participant shared, \textit{``The text view is boring. I~strongly prefer a more visual presentation method.''} Another participant appreciated the actionable insights, stating, \textit{``It's presenting tips without a lot of effort on my part.''}

In summary, participants seem to prefer a combination of the dashboard visuals and the AI chat agent, offering both a visually appealing presentation and the ability to dive deeper into specific insights. Participants also expressed interest in using the data as a tool for having productive conversations with their managers and improving work-life balance. One participant suggested, \textit{``It would be super useful to me, especially as a manager to compare my results or my team's results with others in my workgroup, org, company, etc.''} To further improve the participant experience and the potential of the interactive chat agent, incorporating benchmarking, tracking changes over time, and enhancing the chat agent's capabilities to display visuals could be beneficial.

%% file: 8_discussion.tex
\section{Discussion}
\label{sec:discussion}

Our study employed a two-phase approach to explore productivity agent design: first investigating user preferences through surveys to define an ideal agent persona, then developing and evaluating an AI-based productivity agent against alternative telemetry data presentation modes like dashboards and narratives. The user survey in phase one examined aspects of productivity, communication style, agent traits, personalization, and privacy. Findings showed that participants valued progress on scheduled tasks, resolving unexpected tasks, and assisting colleagues as key productivity indicators. Most participants supported the idea of a productivity agent and believed its personality would significantly impact effectiveness. They sought AI assistance primarily for meeting summaries, documentation, start-of-day briefs, work summaries, and information retrieval.
These results partially align with ~\citet{kimani2019conversational}, who found emails and meetings were primary productivity distractions, and completing scheduled tasks was the top productivity measure. However, an interesting difference emerged: Kimani et al.'s participants prioritized help with reminders, task scheduling, and distraction management, while our participants ranked these lower. This discrepancy might reflect evolving user expectations shaped by media coverage and public discourse about LLM capabilities rather than direct experience. There is a possibility that the hype and discussions about LLM-powered tools in public and online forums might have swayed participants' perceptions, aligning them with tasks where AI is commonly viewed as beneficial. While our study did  not explicitly distinguish between preferences influenced by hands-on LLM use versus societal narratives, this potential influence suggests that user expectations for AI support are shaped by both personal experience and broader technological narratives. Although both studies involved similar job roles (program managers, developers, engineers), other demographic factors might explain these different preferences.
Regarding agent characteristics, participants preferred a balance of casual and formal communication with direct, concise messages—reflecting modern workplace needs for efficient yet relationship-building communication. They favored knowledgeable, transparent assistants over supervisory roles, and most were comfortable sharing basic data for personalization despite some privacy concerns. These preferences informed our productivity agent design in phase two.

Interestingly, approximately 23\% of participants, as detailed in Table~\ref{tbl:productivity_perspectives} (Appendix~\ref{sec:user_survey_responses}), expressed neutrality or did not believe productivity assistants would be beneficial for them. Their concerns centered around several key issues. One participant notably distinguished between overall productivity and motivational challenges, suggesting that while they managed productivity well, they struggled with motivation for unengaging or poorly defined tasks—a challenge they believed was beyond the scope of productivity assistants. Participants also raised concerns about AI reliability, particularly regarding factual accuracy, and worried about over-dependence on AI potentially hampering personal growth and skill development. Some expressed skepticism about AI's practical value in conversations, emphasizing the need for empirical validation of such tools' effectiveness. These insights suggest that while AI agents might excel at providing structural support, they may fall short in addressing intrinsic motivational factors. Understanding these reservations is crucial for improving AI productivity tools' design and increasing their workplace adoption~\cite{10388088, 10.1145/3613904.3642716, 10.1145/3632754.3632939}.

In the second phase of the study, we developed the AI productivity agent based on insights from our initial survey and compared its performance with two alternative methods of presenting telemetry data: a visual dashboard and an AI-based personalized narrative. The interaction data revealed significant differences in user engagement patterns, with users spending notably more time with the AI productivity agent compared to other presentation modes. The AI agent provided more comprehensive responses, with a higher word count per message. This extended engagement duration presents an interesting dimension to our findings. While it could reflect the agent's ability to provide detailed, comprehensive responses requiring more time to digest, it might also be partially attributed to inherent AI response generation delays. Though longer engagement times do not necessarily indicate higher productivity or satisfaction, they suggest significant user interest and willingness to explore AI capabilities in productivity workflows. The depth of interaction and detailed responses point to a potentially richer user experience not present in other tools. Future research should investigate the distinct impacts of response time versus response depth on user experience, engagement, and productivity. The AI agent showed promise in providing tailored insights and recommendations, with most participants agreeing that telemetry data enhanced its ability to provide personalized and empathetic support. Users recognized the agent's demonstration of desired traits like knowledgeability, trustworthiness, and transparency, while also noting its success in avoiding negative characteristics identified in the survey (being overly critical, intrusive, or annoying). Beyond customizing responses, the telemetry data served to foster self-reflection, aligning with previous research showing how such data can enhance reflective practices ~\cite{mcduff2012affectaura, Kocielnik2018}. However, only a few participants found the agent adaptive in its conversation style, possibly due to the study's brief interaction period. Future longitudinal studies could better evaluate the agent's conversational adaptiveness by allowing extended user interactions.

Participants expressed a preference for combining the dashboard's visual appeal and ease of understanding with the chat agent's in-depth, personalized insights. This preference aligns with prior research suggesting that visualizations excel at providing overviews and trend identification, while text summaries better serve targeted information searches \cite{Smith06}. The complementary nature of these approaches—visual data representation for quick comprehension and interactive dialogue for detailed, personalized recommendations—indicates users' desire for a comprehensive productivity tool that leverages both methods' strengths. Future improvements should focus on integrating visualizations into the chat agent's responses and adding benchmarking functionality to allow users to compare their productivity metrics against team or organizational averages. The mixed reception of the personalized narrative method underscores the importance of supporting diverse user preferences through multiple presentation methods or a hybrid approach. In summary, while the interactive chat agent shows promise in providing tailored insights and recommendations, our analysis revealed areas for improvement. For example, we found inconsistencies in the agent's advice, as illustrated in Appendix~\ref{sec:sample_resposne_aiagent} Figure~\ref{fig:sample_chat3}, where the agent contradicts itself on email management strategies—suggesting both specific time slots and even distribution throughout the day. Such contradictions may have influenced participants' perceptions and ratings of the agent, highlighting the need for more refined prompt engineering and improved response generation logic. Future iterations will prioritize enhancing the coherence and consistency of AI recommendations to build better user satisfaction and trust.

Our findings suggest that incorporating design principles like semantic zooming ~\cite{Shneiderman} and progressive disclosure ~\cite{progressivedisclosure} could enhance the effectiveness of productivity tools. Semantic zooming would allow users to explore data at varying levels of detail, facilitating better understanding of relationships between productivity metrics and patterns. Progressive disclosure would present information in layers, starting with essential details and revealing more as needed, aligning with users' preferences for concise yet detailed guidance. Integrating these principles into future AI productivity agents could help strike an optimal balance between high-level overviews and in-depth insights, while maintaining user control over information density. Our study also explored GPT-4's capabilities in transforming productivity data from platforms like Viva Insights into more personalized, interactive advice through an AI chatbot. GPT-4's advanced natural language processing abilities enabled us to create a conversational interface that went beyond simple data visualization, offering tailored recommendations based on individual work patterns. This approach enhanced traditional productivity tools by making insights more actionable for users. While developing the agent, we prioritized user privacy by adhering to privacy-by-design principles, using only high-level telemetry data with explicit user consent. Our research demonstrates how LLMs can enhance user experience while maintaining privacy, contributing to the growing body of work on AI-enhanced productivity tools. Future research should examine different presentation and functionality variations to identify the most effective approaches for enhancing user productivity.

\section{Limitations \& Future Work} 
\label{sec:limitations}
Our study has several limitations worth noting. The sample size and demographic composition may not fully represent the diversity of user preferences and productivity patterns across different professional contexts. The relatively short duration of our study also limited participants' ability to fully adapt to and evaluate the different presentation modes, potentially affecting the long-term validity of our findings. 
While our study provides insights into how different personal informatics tools support productivity-related reflection and action, we acknowledge that we did not measure direct productivity improvements through quantitative metrics such as task completion rates, distraction reduction, or comparative efficiency measures. A longitudinal study examining how each presentation mode influences behavior change and productivity outcomes would be valuable for validating the systems' effectiveness. It is also important to acknowledge how prompt variations can influence GPT-4's outputs. The model's sensitivity to prompt phrasing and structure means that even subtle changes could yield significantly different responses, potentially affecting the consistency and comparability of our findings.

In addition, while we developed prompts through iterative testing, we did not conduct systematic prompt engineering to optimize the balance between accuracy and computational cost. Also of note are the occasional hallucinations exhibited by GPT-4 and its poor ability to perform calculations accurately. As a result, we refrained from asking the agent to generate additional summary information or perform complex calculations, which could have potentially provided more insights into the productivity patterns. While we did not systematically analyze the chat transcripts for hallucinations, it is noteworthy that none of the participants reported experiencing incorrect or hallucinated information in their open-ended survey responses about the agent's performance. In our survey methodology (Appendix~\ref{sec:user_experience}), detailed questions about the interactive agent were followed by queries on the solutions' strengths, weaknesses, and suggestions (Q28, Q29). This sequence could potentially induce a response bias, where participants' views on the agent's performance might have disproportionately influenced their assessment of its strengths and weaknesses. While we aimed for a comprehensive understanding of user experiences and randomized other aspects of the study including Q27 which explicitly asks for the benefits and drawbacks of each presentation method immediately after interacting with that method, future studies may benefit from randomizing the order of these questions to mitigate any potential bias.

The telemetry data extracted from Viva Insights provided valuable information on participant work patterns but did not capture a complete picture. While we focused on macro-level telemetry data such as meeting patterns, email habits, and focus time blocks, we acknowledge that micro-level data (such as moment-to-moment task switches, detailed distraction patterns, or granular activity logs) could potentially enable more fine-grained personalization. Our choice of macro-level telemetry was driven by considerations of privacy, data availability, and practical applicability in enterprise settings. However, this higher-level view might miss productivity patterns that could be captured through more detailed activity tracking. Future research could explore how combining both macro and micro-level telemetry data might enable more sophisticated personalization while carefully balancing the trade-offs between insight granularity and privacy concerns. Furthermore, there might have been external factors or variables, such as changes in workload, organizational structure, or personal circumstances, that may affect productivity and thus are essential to be considered. In addition, our system was designed to extract specific, predefined types of information from telemetry data (such as meeting patterns and email habits), with these categories hardcoded into the prompts. We did this to ensure consistent experimental conditions across participants and facilitate meaningful comparisons between different presentation methods. Since the study primarily aimed to demonstrate the feasibility and effectiveness of integrating GPT-4 with productivity tools, hence this structured approach to data extraction. While this design choice limited the system's flexibility in discovering additional productivity patterns, we believe that with the advancing capabilities of LLMs and robust error handling, future implementations could adopt a more dynamic approach to telemetry data extraction and insight generation.

In light of these limitations, it is essential to approach the study's findings with caution and acknowledge that the results may not be universally applicable. Future research could address these limitations by employing larger and more diverse samples, extending the study duration, incorporating additional measures of productivity, and controlling for potential confounding factors. Finally,  it is important to consider the extent to which the AI productivity agent's design was influenced by the user preferences identified in Phase 1. While there is a possibility that a closer alignment might have affected the comparative results, our design was a conscientious effort to balance user needs with technical feasibility. Future research could explore a more in-depth alignment of these insights, potentially offering a refined understanding of how well the AI agent meets user expectations. This ongoing process highlights the iterative nature of developing user-centric AI tools.

\section{Ethical Considerations}
\label{sec:ethical_considerations}
We adhered to strict ethical standards throughout our study to protect participant rights and privacy. All participants provided informed consent after receiving a comprehensive overview of the study's objectives, procedures, potential benefits, and risks. Anonymity and confidentiality were maintained by using unique identification codes for participant data, securely storing all information on encrypted servers, and limiting access to authorized research team members. Our data security approach extends beyond the study period through strict data minimization practices. Following study completion, all participant contact information used for payment and study coordination was permanently deleted from our systems. The remaining anonymized research data is stored using enterprise-grade encryption and follows the organization's established data retention policies, which include regular security assessments and updates to protect against emerging vulnerabilities. This approach ensures participant information remains protected long-term.

The use of workplace telemetry data, particularly when combined with generative AI technologies, raises significant concerns around user privacy, autonomy, and workplace surveillance ~\cite{10.1145/3663384.3663401, 9597452, 10.1145/3544548.3581376}. These concerns are especially pertinent as organizations increasingly adopt sophisticated tracking and analytics tools that could potentially enable intrusive monitoring practices. During the telemetry data collection process, we implemented several specific privacy-preserving measures. Our scraper tool's interface provided participants complete visibility into the data being collected, and importantly, participants could review and modify their telemetry data before submission. We adhered to the principle of data minimization by collecting only pre-defined, high-level metrics without accessing detailed communication content, and limiting historical data to the previous month to minimize data exposure. All collected data was immediately encrypted and stored in a secure environment. This approach to telemetry data personalization was validated by our user survey findings, where 73.2\% of participants expressed comfort in sharing high-level telemetry data for enhancing productivity guidance.

Beyond data collection considerations, the design of AI-based productivity systems presents additional ethical challenges. We deliberately chose to work with high-level aggregated telemetry data rather than granular activity tracking, striking a balance between providing meaningful insights and respecting user privacy. Our implementation ensures that the system serves as a supportive tool for individual productivity improvement rather than a monitoring mechanism for management oversight. The use of GPT-4 in processing workplace telemetry data introduces additional ethical considerations specific to our context. We addressed these through several design choices: utilizing the enterprise version of GPT-4 purely as an inference engine without data retention, processing only aggregated productivity metrics rather than detailed personal communications or schedules, and ensuring that all recommendations are suggestive rather than prescriptive. This approach maintains user autonomy while providing productivity support.
Furthermore, we considered the impact of AI-driven productivity recommendations on work-life balance and employee well-being. While our system makes suggestions based on telemetry data, we designed it to avoid inadvertently promoting overwork or unhealthy work patterns. These ethical considerations informed both our technical implementation and evaluation approach, where we prioritized understanding user preferences and comfort levels with different types of productivity guidance. Our goal was to develop a system that enhances workplace productivity while respecting user privacy, autonomy, and well-being.

%% file: 9_conclusion.tex
\section{Conclusion}
\label{sec:conclusion}
This paper presents a comprehensive study focused on understanding user preferences for productivity agents and designing personalized solutions to cater to these preferences. By conducting a user survey and analyzing the findings, we were able to identify key features and behaviors that contribute to user satisfaction and comfort when interacting with a productivity agent. Subsequently, we developed an AI-based productivity agent and compared its performance with alternative modes of telemetry data presentation, such as dashboard visualization and personalized narratives. The results from this study provide useful insights into the factors that enhance productivity and user experience, as well as areas for improvement in the design and development of personalized productivity agents. By incorporating the lessons learned from this study, future work can focus on refining and optimizing productivity-enhancing tools and solutions, ultimately leading to improved efficiency and user-centric experiences for information workers.

%% file: appendix.tex
\appendix
\clearpage
\section{User survey questions}
\label{sec:user_survey}
The table below lists the questions we asked participants during the first phase of the study i.e., user survey.
\begin{table*}[!htbp]
\caption{User survey questions.}
\smaller
\begin{tabular}{@{}lllll@{}}
\textbf{Category}                      \label{tbl:usersurvey}                                                                                                                      & \textbf{Questions} \\ \bottomrule
{\cellcolor[HTML]{EFEFEF} Productivity Perspectives} & (1) What makes you feel productive at the end of your workday?\\  & (2) Do you believe a productivity agent would be beneficial for you in managing your personal\\ & \thinspace \thinspace  \thinspace \thinspace \thinspace \thinspace \thinspace \thinspace \thinspace \thinspace and professional tasks, improving efficiency, and enhancing your overall productivity?\\  & (3) Which productivity areas do you struggle with or would you like an AI agent to provide \\ & \thinspace \thinspace  \thinspace \thinspace \thinspace \thinspace \thinspace \thinspace \thinspace \thinspace guidance and suggestions for, in order to help you improve your productivity and maintain \\ & \thinspace \thinspace  \thinspace \thinspace \thinspace \thinspace \thinspace \thinspace \thinspace \thinspace better control over your work? *\\ & (4) How do you think the personality of your productivity agent would impact its effectiveness \\ & \thinspace \thinspace  \thinspace \thinspace \thinspace \thinspace \thinspace \thinspace \thinspace \thinspace as a productivity tool? \\
{\cellcolor[HTML]{EFEFEF} Communication Style}  & (5) What is your preference regarding the communication tone of the agent?\\
& (6) What kind of language or tone would make you feel most comfortable opening up to your \\ & \thinspace \thinspace  \thinspace \thinspace \thinspace \thinspace \thinspace \thinspace \thinspace \thinspace  productivity agent about any concerns related to productivity, in order to \\ & \thinspace \thinspace  \thinspace \thinspace \thinspace \thinspace \thinspace \thinspace \thinspace \thinspace foster a supportive and trusting environment that encourages open communication \\ & \thinspace \thinspace  \thinspace \thinspace \thinspace \thinspace \thinspace \thinspace \thinspace \thinspace and effective problem-solving? \\
& (7) What style of messages would you prefer?\\
{\cellcolor[HTML]{EFEFEF} Approach \& Personality Traits} & (8) In what manner would you prefer your productivity agent to provide assistance for \\ & \thinspace \thinspace  \thinspace \thinspace \thinspace \thinspace \thinspace \thinspace \thinspace \thinspace enhancing your productivity?\\
& (9) Are there any specific personality traits you would like the AI productivity agent to possess, \\ & \thinspace \thinspace  \thinspace \thinspace \thinspace \thinspace \thinspace \thinspace \thinspace \thinspace which would make you feel more comfortable interacting with it? *\\
& (10) Which social role would you prefer your productivity agent to have in order to foster a positive \\ & \thinspace \thinspace  \thinspace \thinspace \thinspace \thinspace \thinspace \thinspace \thinspace \thinspace and effective working relationship? *\\
{\cellcolor[HTML]{EFEFEF} Personalization \& Privacy }& (11) Would you like your productivity agent to offer personalized recommendations based on your \\ & \thinspace \thinspace  \thinspace \thinspace \thinspace \thinspace \thinspace \thinspace \thinspace \thinspace work habits and preferences?  \\
& (12) Would you feel comfortable sharing personal information with your productivity agent\\ & \thinspace \thinspace  \thinspace \thinspace \thinspace \thinspace \thinspace \thinspace \thinspace \thinspace in order to improve its recommendations?\\
& (13) Would you be comfortable sharing high level telemetry data (such as the number of emails sent \\ & \thinspace \thinspace  \thinspace \thinspace \thinspace \thinspace \thinspace \thinspace \thinspace \thinspace or meetings attended) with your productivity agent to receive tailored tips and guidance for\\ & \thinspace \thinspace  \thinspace \thinspace \thinspace \thinspace \thinspace \thinspace \thinspace \thinspace enhancing productivity?\\
{\cellcolor[HTML]{EFEFEF} Negative traits}& (14) In your opinion, what are some potential negative personality traits that should be avoided in a \\ & \thinspace \thinspace  \thinspace \thinspace \thinspace \thinspace \thinspace \thinspace \thinspace \thinspace productivity agent to ensure a positive, effective, and supportive user experience? \\
\bottomrule
\end{tabular}
\end{table*}

\clearpage

\section{User Experience and Preferences}
\label{sec:user_experience}
The table below lists the questions we asked participants during the second phase of the study for evaluation.

\begin{table*}[!htbp]
\caption{Evaluation questions.}
\smaller
\begin{tabular}{@{}lllll@{}}
\textbf{Category}                      \label{tbl:usersurvey}                                                                                                                      & \textbf{Questions} \\ \bottomrule
{\cellcolor[HTML]{EFEFEF} Overall Experience} & (1) Please rate your overall experience with the data presentation method\\  
{\cellcolor[HTML]{EFEFEF} Specific Experiences}  
& Considering specific aspects of the data presentation method you interacted with, please rate \\ & \thinspace \thinspace \thinspace \thinspace \thinspace \thinspace \thinspace \thinspace \thinspace \thinspace your experience in the following categories: \\ & \thinspace \thinspace  \thinspace \thinspace \thinspace \thinspace \thinspace \thinspace \thinspace \thinspace (2) It was easy to explore the presentation method and understand the presented data\\  & \thinspace \thinspace  \thinspace \thinspace \thinspace \thinspace \thinspace \thinspace \thinspace \thinspace  (3) I enjoyed interacting with the presentation method \\ & \thinspace \thinspace  \thinspace \thinspace \thinspace \thinspace \thinspace \thinspace \thinspace \thinspace  (4) I was engaged during data exploration \\ & \thinspace \thinspace  \thinspace \thinspace \thinspace \thinspace \thinspace \thinspace \thinspace \thinspace  (5) The data presentation was intuitive \\ & \thinspace \thinspace  \thinspace \thinspace \thinspace \thinspace \thinspace \thinspace \thinspace \thinspace (6) I would use the presentation method to explore my telemetry data in real-life situation \\ &  \thinspace \thinspace  \thinspace \thinspace \thinspace \thinspace \thinspace \thinspace \thinspace \thinspace (7) I feel that the data presented is useful enough to incorporate into my daily life, such as \\ & \thinspace \thinspace  \thinspace \thinspace \thinspace \thinspace \thinspace \thinspace \thinspace \thinspace  \thinspace \thinspace \thinspace  \thinspace \thinspace \thinspace  \thinspace \thinspace \thinspace goal setting and calendar planning, for making informed behavioral decisions\\ &  \thinspace \thinspace  \thinspace \thinspace \thinspace \thinspace \thinspace \thinspace \thinspace \thinspace (8) I trust the accuracy of information presented \\
{\cellcolor[HTML]{EFEFEF}Anticipated Usage Frequency} & (9) How often do you anticipate using the data presentation method?\\
{\cellcolor[HTML]{EFEFEF} Preference Ranking} & (10) Please rank the three methods of data presentation (dashboard, personalized narrative, and \\ & \thinspace \thinspace  \thinspace \thinspace \thinspace \thinspace \thinspace \thinspace \thinspace \thinspace interactive chat agent) in order of preference, with 1 being your most preferred and 3 being\\ & \thinspace \thinspace  \thinspace \thinspace \thinspace \thinspace \thinspace \thinspace \thinspace \thinspace  your least preferred. \\
{\cellcolor[HTML]{EFEFEF} Interactive Agent Capabilities}  
& Below is a list of statements about your interaction with the interactive chat agent. Please read \\ & \thinspace \thinspace \thinspace \thinspace \thinspace \thinspace \thinspace \thinspace \thinspace \thinspace each statement carefully and rate your agreement with the statement. There are no right or \\ & \thinspace \thinspace \thinspace \thinspace \thinspace \thinspace \thinspace \thinspace \thinspace \thinspace wrong answers or trick statements. Please answer each question as honestly as you can. \\ & \thinspace \thinspace \thinspace \thinspace \thinspace \thinspace \thinspace \thinspace \thinspace \thinspace If you think you were not able to get enough information to get a feel of a particular statement,\\ & \thinspace \thinspace \thinspace \thinspace \thinspace \thinspace \thinspace \thinspace \thinspace \thinspace please select "Non-applicable" \\ & \thinspace \thinspace  \thinspace \thinspace \thinspace \thinspace \thinspace \thinspace \thinspace \thinspace (11) The agent provided useful tips and strategies for improving productivity\\  & \thinspace \thinspace  \thinspace \thinspace \thinspace \thinspace \thinspace \thinspace \thinspace \thinspace  (12) The agent was able to understand my productivity challenges\\ & \thinspace \thinspace  \thinspace \thinspace \thinspace \thinspace \thinspace \thinspace \thinspace \thinspace  (13) The agent's empathy and tone helped me feel more comfortable discussing my productivity\\ & \thinspace \thinspace  \thinspace \thinspace \thinspace \thinspace \thinspace \thinspace \thinspace \thinspace \thinspace \thinspace \thinspace \thinspace \thinspace \thinspace \thinspace \thinspace \thinspace \thinspace \thinspace \thinspace \thinspace concerns \\ & \thinspace \thinspace  \thinspace \thinspace \thinspace \thinspace \thinspace \thinspace \thinspace  (14) The agent personalized its suggestions based on my situation and needs\\ & \thinspace \thinspace  \thinspace \thinspace \thinspace \thinspace \thinspace \thinspace \thinspace \thinspace (15) The agent provided practical solutions to overcome productivity obstacles \\ &  \thinspace \thinspace  \thinspace \thinspace \thinspace \thinspace \thinspace \thinspace \thinspace \thinspace (16) The agent used a professional and direct language and tone \\ &  \thinspace \thinspace  \thinspace \thinspace \thinspace \thinspace \thinspace \thinspace \thinspace \thinspace (17) The agent proactively suggested ways to improve my productivity \\ &  \thinspace \thinspace  \thinspace \thinspace \thinspace \thinspace \thinspace \thinspace \thinspace \thinspace (18) The agent was annoying, aggressive or pushy \\ &  \thinspace \thinspace  \thinspace \thinspace \thinspace \thinspace \thinspace \thinspace \thinspace \thinspace (19) The agent was being too intrusive \\ &  \thinspace \thinspace  \thinspace \thinspace \thinspace \thinspace \thinspace \thinspace \thinspace \thinspace (20) The agent was overly critical \\ &  \thinspace \thinspace  \thinspace \thinspace \thinspace \thinspace \thinspace \thinspace \thinspace \thinspace (21) The agent seems to be able to easily work out what I might want to talk about \\ &  \thinspace \thinspace  \thinspace \thinspace \thinspace \thinspace \thinspace \thinspace \thinspace \thinspace (22) The agent seems to have a difficult time seeing things from my point of view \\
 &  \thinspace \thinspace  \thinspace \thinspace \thinspace \thinspace \thinspace \thinspace \thinspace \thinspace (23) The agent seems to vary its conversational style to accommodate my mood or disposition \\
 &  \thinspace \thinspace  \thinspace \thinspace \thinspace \thinspace \thinspace \thinspace \thinspace \thinspace (24) The agent seems to follow what I say and accurately reflects its understanding to me\\
 &  \thinspace \thinspace  \thinspace \thinspace \thinspace \thinspace \thinspace \thinspace \thinspace \thinspace (25) The agent demonstrated the traits of being knowledgeable, trustworthy, transparent,\\ & \thinspace \thinspace  \thinspace \thinspace \thinspace \thinspace \thinspace \thinspace \thinspace \thinspace \thinspace \thinspace \thinspace \thinspace \thinspace \thinspace \thinspace \thinspace \thinspace \thinspace \thinspace \thinspace \thinspace  and responsive \\
 &  \thinspace \thinspace  \thinspace \thinspace \thinspace \thinspace \thinspace \thinspace \thinspace \thinspace (26) The agent's use of telemetry data enhanced its ability to provide personalized and \\ & \thinspace \thinspace  \thinspace \thinspace \thinspace \thinspace \thinspace \thinspace \thinspace \thinspace \thinspace \thinspace \thinspace \thinspace \thinspace \thinspace \thinspace \thinspace \thinspace \thinspace \thinspace \thinspace \thinspace  empathetic support \\
{\cellcolor[HTML]{EFEFEF} Strengths, Weaknesses \& Suggestions }& (27) After interacting with the data presentation method, what do you perceive as the benefits and \\ & \thinspace \thinspace  \thinspace \thinspace \thinspace \thinspace \thinspace \thinspace \thinspace \thinspace \thinspace \thinspace \thinspace \thinspace \thinspace \thinspace \thinspace \thinspace \thinspace \thinspace \thinspace \thinspace \thinspace  drawbacks of using this approach? Please share your thoughts on both the pros and cons in \\ & \thinspace \thinspace  \thinspace \thinspace \thinspace \thinspace \thinspace \thinspace \thinspace \thinspace \thinspace \thinspace \thinspace \thinspace \thinspace \thinspace \thinspace \thinspace \thinspace \thinspace \thinspace \thinspace \thinspace   an open-ended response. Please do not include any personally identifiable information \\ & \thinspace \thinspace  \thinspace \thinspace \thinspace \thinspace \thinspace \thinspace \thinspace \thinspace \thinspace \thinspace \thinspace \thinspace \thinspace \thinspace \thinspace \thinspace \thinspace \thinspace \thinspace \thinspace \thinspace  in your responses.  \\
& (28) Were there any specific features or aspects of each data presentation method that you particularly \\ & \thinspace \thinspace  \thinspace \thinspace \thinspace \thinspace \thinspace \thinspace \thinspace \thinspace \thinspace \thinspace \thinspace \thinspace \thinspace \thinspace \thinspace \thinspace \thinspace \thinspace \thinspace \thinspace \thinspace liked or disliked?  Please provide details. Please do not include any personally identifiable  \\ & \thinspace \thinspace  \thinspace \thinspace \thinspace \thinspace \thinspace \thinspace \thinspace \thinspace \thinspace \thinspace \thinspace \thinspace \thinspace \thinspace \thinspace \thinspace \thinspace \thinspace \thinspace \thinspace \thinspace  information in your responses.\\
& (29) Please share any thoughts or suggestions you might have about the study, the survey questions, \\ & \thinspace \thinspace  \thinspace \thinspace \thinspace \thinspace \thinspace \thinspace \thinspace \thinspace \thinspace \thinspace \thinspace \thinspace \thinspace \thinspace \thinspace \thinspace \thinspace \thinspace \thinspace \thinspace \thinspace or the bot's performance. For example, what improvements or additional features would you \\ & \thinspace \thinspace  \thinspace \thinspace \thinspace \thinspace \thinspace \thinspace \thinspace \thinspace \thinspace \thinspace \thinspace \thinspace \thinspace \thinspace \thinspace \thinspace \thinspace \thinspace \thinspace \thinspace \thinspace like to see? Please do not include any personally identifiable information in your responses. \\

\bottomrule
\end{tabular}

\end{table*}

\section{Prompts for Extracting Data}
\label{sec:prompts_extracting_data}
\begin{tcolorbox}
\textbf{System Prompt: }You are an agent that parses useful information from given HTML source. The user will give you an HTML source and from that you should identify the information the user requested and send it to them in proper format. Do not make stuff up if you cannot find it in the HTML source. If a thing the user asked for doesn't exist, do not include it in your response. \newline 
\newline
\textbf{User Prompt for Page 1: }Extract the organized meetings, invited meetings and meeting habits (along with \%) and list them. If you cannot find these information, do not mention anything about it in the response. DO NOT LIST ANY NAMES OF INDIVIDUALS. List the information in this format: `Meetings organized in the past four weeks: X,  Meetings invited to in the past four weeks: X, Meeting habits based on meetings you attended: - Advanced notice: X\%, - High attendance: X\%, - No overlap with other meetings: X\%, - Ended on time: X\%,- Joined on time: X\% - During working hours: X\% - Meeting was $\leq$ 1 hour: X\% - RSVP'd to invite: X\% - You didn't multitask: X\% - Added a teams link: X\%'\newline
\textbf{User Prompt for Page 2: }Extract communication habits and hourly breakdown. If you cannot find communication habits or hourly breakdown, do not mention anything about it in the response. DO NOT LIST ANY NAMES OF INDIVIDUALS. List the information in this format: `Communication habits: - Emails sent: X - Emails read: X - Chats and calls: X Hourly breakdown of communications (past 4 weeks): * list the hourly range data in the format:* - 10 AM to 11 AM: X emails sent, X emails read, X chats/calls' 
\newline
\textbf{User Prompt for Page 3: }Extract focus plan and disconnect with quiet time. If you cannot find both focus plan and disconnect with quiet time both, then just respond EXACTLY and ONLY with this: `No data on focus time'. DO NOT LIST ANY NAMES OF INDIVIDUALS. If you find disconnect with quiet time, and the sentence reads like this: `X out of Y days have been interrupted by emails, chats, and meetings after work hours.', then list the information in this format: `Disconnect with quiet time: X out of Y days have been interrupted by emails, chats, and meetings after work hours.'. However, if the disconnect with quiet time sentence reads like this: `X out of Y days without quiet time interruptions', then first subtract X from Y so you have Z, then list the information in this format: `Disconnect with quiet time: Z out of Y days have been interrupted by emails, chats, and meetings after work hours.'. If you find focus plan, list the information in the following format: `Focus Plan: You kept X hours and X minutes of focus time in Y. You have X hours of focus time reserved next week.'. IMPORTANT: If you do not find focus time but find disconnect with quiet time, do not mention anything about focus time. If you do not find disconnect with quiet time but find focus time, do not mention anything about disconnect with quiet time. If you don't find both focus time and disconnect with quiet time, then do not respond anything else other than: `No data on focus time'. 
\newline
\textbf{User Prompt for Page 4:} Extract communication habit (along with \%) and collaboration habits (time spent collaborating and number of collaborators). Extract in dictionary format, using this template: \{ \texttt{`available\_to\_focus': value, `meetings': value, `chats': value, `emails': value, `number\_of\_collaborators': value, `collab\_within\_working': value, `collab\_outside\_working': value} \}. If you cannot find a certain value, put null as value. DO NOT LIST ANY NAMES OF INDIVIDUALS.
\end{tcolorbox}

\section{Prompt for Transforming Data}
\label{sec:prompts_transfomring_data}
\begin{tcolorbox}
\textbf{User Prompt for transforming to JSON: }Here is some data: [USER'S TELEMETRY DATA]. Convert the above data to the JSON format shown below. Do not return anything other than the JSON. Use relevant numbers you can find in the above text for each JSON entry. If any of the data is missing,  enter null in its place.

Here is a sample input: [SAMPLE INPUT OF THE TELEMETRY DATA]

And here it is converted in JSON format:
\begin{verbatim}  
'Meetings': {
    'organized': 7,'invited': 43,
    'habits': {
      'Advanced notice': 29, 'High attendance': 70,'No overlap with other meetings': 74,
      ...
    } 
    },'CommunicationHabits': {
    'emailsSent': 15032, 
    'emailsRead': 348, 
    'chatsAndCalls': 182,
   'hourlyBreakdown': {
  '10:00 AM–11:00 AM': {'sent': 2, 'read': 20, 'chats': 16},
  ....
} 
}, 'DisconnectWithQuietTime': {
    'daysInterrupted': 19, 'totalDays': 28
  },
  'FocusPlan': {
    'focusTimeThisMonth': '35 hr 55 mins', 'focusTimeMonth': 'July', 
    ...
  }, 'Collaborations': {
    'pieChart': {
      'Available to focus': 83.47,
      'Meetings': 9.56,
      'Chats': 3.39,
      'Emails': 3.59
    },
    'withinWorkingHours': 46,'outsideWorkingHours': 7, 'numberOfCollaborators': 58
  }
  }
\end{verbatim}  

Follow the same format as the JSON above. Use the exact same keys. The values can be different, but the keys should be the same. If you cannot find something, you should use null. DO NOT RETURN ANYTHING OTHER THAN JSON.
\end{tcolorbox}

\section{Prompt for Personalized Narrative and Interactive Chat Agent}
\label{sec:prompts_agents}
\begin{tcolorbox}
\textbf{System Prompt for Interactive Agent: }I am an AI-powered productivity and well-being expert designed to help users achieve an optimal work-life balance and improve their productivity. Professional, direct, and knowledgeable, I understand the intricacies of modern work environments and offer personalized strategies to enhance productivity while focusing on user's overall well-being. I have access to user's telemetry data. For every query from the user, I try to identify if there is any relevant telemetry data, and provide personalized recommendations tailored to user's unique work patterns and goals. As a trustworthy and responsive expert assistant, I use a directive yet supportive approach, proactively suggesting improvements without being pushy or intrusive. The language and tone I use are context-aware, ensuring clear and concise communication without being overly chatty or verbose. Empathy and transparency are at the core of my interactions, avoiding any judgmental or patronizing attitudes. I am committed to providing accurate and reliable information to help users reduce stress, improve well-being, and maximize their productivity potential, all while making the most of the telemetry data. I continuously learn and adapt to user's evolving needs. When referring to the telemetry data in my response, I use variations such as, 'according to the telemetry data', 'as per the telemetry data', 'considering the telemetry data', 'as informed by the telemetry data', 'taking into account the telemetry data', 'drawing upon the telemetry data', and so on. I do not use any markdown elements such as stars/asterisks i.e. * or ** in my response to bold, underline, italicize or emphasize texts. The following is the user's telemetry data: [USER'S TELEMETRY DATA]\newline 
\textbf{System Prompt for Personalized Narrative: } I am an expert AI assistant, committed to providing accurate and efficient personalized narrative based on user's telemetry data. I examine the given telemetry data, determine important connections, and present an extensive narrative with practical recommendations in a proficient way. The personalized narrative I present examines the telemetry data, identifies key findings, trends and significant patterns. I follow the following format while presenting the extensive, personalized narrative: the first paragraph about meetings, second paragraph on communication habits, the third paragraph on focus time, quiet time and collaborations. In the final paragraph I offer actionable insights based on my findings and analysis in bullet points, calling out the telemetry data when relevant. I ensure that no critical information is left out. I have the ability to understand the context in which the telemetry data is being used to provide more targeted and relevant insights. I create narratives that are engaging and interesting to the user, encouraging them to explore the insights further and take appropriate actions. When referring to the telemetry data in my response, I use variations such as, 'based on the telemetry data', 'according to the telemetry data', 'as per the telemetry data', 'considering the telemetry data', 'as informed by the telemetry data', 'taking into account the telemetry data', 'drawing upon the telemetry data'. I present the narrative in a clear and concise manner, making it easy for the user to understand and interpret the findings. 

I do not mention the user's name or my name in the response. I ensure that the data analysis and interpretations I make are accurate. The narrative I make are easily understandable by the user. I avoid using overly technical language or jargon that could be confusing. I present the information in a concise and efficient manner, avoiding repetition of points that have already been made. I maintain objectivity in my analysis and avoid introducing any personal biases or assumptions.\newline
\end{tcolorbox}

\section{Demographic distribution of user survey.}
\label{sec:demographics_firstphase}
\begin{table}[ht!]
\small
\caption{Demographics of the participants. The table below lists the demographic composition of the participants in our user survey.}
\begin{tabular}{@{}lllll@{}}
\textbf{Category}                                                                                                                                            & \textbf{Count}   & \textbf{Percentage}\\ \bottomrule
\multicolumn{3}{l}{\cellcolor[HTML]{EFEFEF}\textit{Gender}}                                                                        \\
Woman & 109 & 30.0\%   \\
Man   & 244 & 67.2\%    \\
Non-binary/gender diverse & 4 & 1.1\%\\ 
Prefer not to answer & 6 & 1.7\% \\ 
\multicolumn{3}{l}{\cellcolor[HTML]{EFEFEF}\textit{Age}}    \\
18-25 & 20 & 5.5\% \\
26-35 & 96 & 26.4\% \\
36-45 & 100 & 27.5\%  \\
46-55 & 97 & 26.7\%  \\
56-65 & 37 & 10.2\% \\
66 and above & 5 & 1.4\% \\
Prefer not to say & 8 & 2.2\% \\
\multicolumn{3}{l}{\cellcolor[HTML]{EFEFEF}\textit{Job Role}}    \\
Software development/engineering & 168 & 46.3\% \\
Product management & 71 & 19.6\%  \\
Administrative/operations & 20 & 5.5\% \\
Data science/analytics & 14 & 3.9\% \\
Customer Support & 11  & 3.0\% \\
IT/infrastructure & 11 & 3.0\% \\
Other & 65 & 17.9\% \\
Prefer not to say & 3 & 0.8\%\\
\bottomrule
\end{tabular}
  \label{tab:demographics_firstphase}
\end{table}

\section{Demographic distribution of second phase of the study.}
\label{sec:demographics_secondphase}
\begin{table}[ht!]
\small
\caption{Demographics of the participants. The table below lists the demographic composition of the participants in the second phase of our study.}
\begin{tabular}{@{}lllll@{}}
\textbf{Category}                                                                                                                                            & \textbf{Count}   & \textbf{Percentage}\\ \bottomrule
\multicolumn{3}{l}{\cellcolor[HTML]{EFEFEF}\textit{Gender}}                                                                        \\
Woman & 16 & 40.0\%   \\
Man   & 24 & 60.0\%    \\
\multicolumn{3}{l}{\cellcolor[HTML]{EFEFEF}\textit{Age}}    \\
18-25 & 3 & 7.5\% \\
26-35 & 12 & 30.0\% \\
36-45 & 7 & 17.5\%  \\
46-55 & 13 & 32.5\%  \\
56-65 & 5 & 12.5\% \\
\multicolumn{3}{l}{\cellcolor[HTML]{EFEFEF}\textit{Job Role}}    \\
Software development/engineering & 4 & 10.0\% \\
Product management & 21 & 52.5\%  \\
Administrative/operations & 3 & 7.5\% \\
Data science/analytics & 3 & 7.5\% \\
Customer Support & 2  & 5.0\% \\
IT/infrastructure & 1 & 2.5\% \\
Other & 6 & 15.0\% \\
\bottomrule
\end{tabular}
  \label{tab:demographics_secondphase}
\end{table}

\section{User survey responses}
\label{sec:user_survey_responses}
\begin{table*}[h!t!]  
\caption{Productivity Perspectives}   
\smaller
\begin{tabular}{@{}p{0.75\linewidth}l@{}}  
\textbf{Question \& Options} & \textbf{Count} \\ \midrule  
\multicolumn{2}{l}{\cellcolor[HTML]{EFEFEF}\begin{minipage}{0.75\linewidth}\textit{What makes you feel productive at the end of your workday? *(select all that apply) }\end{minipage}} \\  
Making progress or completing scheduled tasks & 347 (95.6\%)   \\
Solving novel, unexpected tasks  & 278 (76.6\%) \\
Helping a colleague accomplish a task  & 262 (72.2\%)\\
Learning new techniques or information & 250 (68.9\%)  \\
Staying up-to-date with messaging and emails  & 153 (42.1\%)\\
Effectively managing and participating in meetings  & 149 (41.0\%)\\
Establishing new connections or expanding my professional network  & 111 (30.6\%)\\
Other & 14 (3.9\%) \\ \midrule  
\multicolumn{2}{l}{\cellcolor[HTML]{EFEFEF}\begin{minipage}{0.75\linewidth}\textit{Do you believe a productivity agent would be beneficial for you in managing your personal and professional tasks, improving efficiency, and enhancing your overall productivity?}\end{minipage}} \\  
Yes, definitely & 142 (39.1\%) \\  
Yes, somewhat & 138 (38.0\%) \\  
Neutral & 57 (15.7\%) \\  
No, not really & 17 (4.7\%) \\  
No, not at all & 9 (2.5\%) \\ \midrule
\multicolumn{2}{l}{\cellcolor[HTML]{EFEFEF}\begin{minipage}{0.75\linewidth}\textit{How do you think the personality of your productivity agent would impact its effectiveness as a productivity tool? }\end{minipage}} \\  
Moderate impact & 117 (41.8\%)   \\
Minimal impact & 73 (26.1\%) \\
Significant impact & 54 (19.3\%)\\
Unsure & 27 (9.6\%)  \\
Other & 9 (3.2\%) \\
\bottomrule  
\end{tabular}  
\label{tbl:productivity_perspectives}  
\end{table*}

\begin{table*}[ht!]  
\caption{Communication Style}   
\smaller
\begin{tabular}{@{}p{0.75\linewidth}l@{}}  
\textbf{Question \& Options} & \textbf{Count} \\ \midrule  
\multicolumn{2}{l}{\cellcolor[HTML]{EFEFEF}\begin{minipage}{0.75\linewidth}\textit{What is your preference regarding the communication tone of the agent?}\end{minipage}} \\  
Casual and friendly & 46 (16.4\%)   \\
Formal and professional  & 23 (8.2\%) \\
It depends on the context & 95 (33.9\%)\\
A balance of both & 99 (35.4\%)  \\
No strong preference & 17 (6.1\%)\\ \midrule  
\multicolumn{2}{l}{\cellcolor[HTML]{EFEFEF}\begin{minipage}{0.75\linewidth}\textit{What kind of language or tone would make you feel most comfortable opening up to your productivity agent about any concerns related to productivity, in order to foster a supportive and trusting environment that encourages open communication and effective problem-solving?}\end{minipage}} \\  
Professional and respectful & 71 (25.4\%) \\  
Direct and straightforward & 69 (24.6\%) \\  
Adaptive & 55 (19.6\%) \\  
Friendly and warm & 50 (17.9\%) \\  
Engaging and motivational & 19 (6.8\%) \\  
Unsure & 12 (4.3\%) \\  
Other & 4 (1.4\%) \\ \midrule
\multicolumn{2}{l}{\cellcolor[HTML]{EFEFEF}\begin{minipage}{0.75\linewidth}\textit{What style of messages would you prefer?}\end{minipage}} \\  
Direct and to the point & 99 (35.4\%)   \\
Chatty and conversational & 12 (4.3\%) \\
A balance of both & 86 (30.7\%)\\
It depends on the context & 80 (28.6\%)  \\
No strong preference & 3 (1.1\%) \\
\bottomrule  
\end{tabular}  
\label{tbl:communication_style}  
\end{table*}  

\begin{table*}[ht!]  
\caption{Approach \& Personality Traits}   
\smaller
\begin{tabular}{@{}p{0.75\linewidth}l@{}}  
\textbf{Question \& Options} & \textbf{Count} \\ \midrule  
\multicolumn{2}{l}{\cellcolor[HTML]{EFEFEF}\begin{minipage}{0.75\linewidth}\textit{Would you like your productivity agent to offer personalized recommendations based on your work habits and preferences?}\end{minipage}} \\  
Moderate Personalization & 109 (38.9\%)   \\
High Personalization  & 100 (35.7\%) \\
Minimal Personalization & 34 (12.1\%)\\
No personalization & 15 (5.4\%)\\
Unsure & 22 (7.9\%)  \\ \midrule  
\multicolumn{2}{l}{\cellcolor[HTML]{EFEFEF}\begin{minipage}{0.75\linewidth}\textit{Would you feel comfortable sharing personal information with your productivity agent in order to improve its recommendations?}\end{minipage}} \\  
Yes, completely comfortable & 46 (16.4\%)   \\
Somewhat comfortable & 106 (37.9\%) \\
Neutral & 61 (21.8\%)\\
Uncomfortable & 58 (20.7\%)  \\
Unsure & 9 (3.2\%)\\\midrule
\multicolumn{2}{l}{\cellcolor[HTML]{EFEFEF}\begin{minipage}{0.75\linewidth}\textit{Would you be comfortable sharing high level telemetry data (such as the number of emails sent or meetings attended) with your productivity agent to receive tailored tips and guidance for enhancing productivity?}\end{minipage}} \\  
Yes, completely comfortable & 121 (43.2\%)   \\
Somewhat comfortable & 84 (30.0\%) \\
Neutral & 36 (12.9\%)\\
Uncomfortable & 31 (11.1\%)  \\
Unsure & 8 (2.9\%)\\
\bottomrule  
\end{tabular}  
\label{tbl:communication_style}  
\end{table*}

\begin{table*}[ht!]  
\caption{Personalization \& Privacy}   
\smaller
\begin{tabular}{@{}p{0.75\linewidth}l@{}}  
\textbf{Question \& Options} & \textbf{Count} \\ \midrule  
\multicolumn{2}{l}{\cellcolor[HTML]{EFEFEF}\begin{minipage}{0.75\linewidth}\textit{In what manner would you prefer your productivity agent to provide assistance for enhancing your productivity?}\end{minipage}} \\  
Proactive suggestions & 169 (60.4\%)   \\
Only specific requests  & 77 (27.5\%) \\
Not at all & 7 (2.5\%)\\
Unsure & 19 (6.8\%)  \\
Other & 8 (2.9\%)\\ \midrule  
\multicolumn{2}{l}{\cellcolor[HTML]{EFEFEF}\begin{minipage}{0.75\linewidth}\textit{Are there any specific personality traits you would like the AI productivity agent to possess, which would make you feel more comfortable interacting with it? *(select all that apply)}\end{minipage}} \\  
Knowledgeable & 249 (88.9\%) \\  
Trustworthy & 210 (75.0\%) \\  
Transparent & 191 (68.2\%) \\  
Responsive & 184 (65.7\%) \\  
Adaptable & 172 (61.4\%) \\  
Professional & 131 (46.8\%) \\  
Friendly & 104 (37.1\%) \\ 
Humor & 92 (32.9\%) \\  
Empathetic & 81 (28.9\%) \\
No strong preference & 4 (1.4\%) \\  
Other & 13 (4.6\%) \\\midrule
\multicolumn{2}{l}{\cellcolor[HTML]{EFEFEF}\begin{minipage}{0.75\linewidth}\textit{Which social role would you prefer your productivity agent to have in order to foster a positive and effective working relationship? *(select all that apply)}\end{minipage}} \\  
Assistant & 214 (76.4\%)   \\
Expert & 157 (56.1\%) \\
Coworker & 96 (34.3\%)\\
Mentor & 93 (33.2\%)  \\
Friend & 65 (23.2\%) \\
Supervisor & 13 (4.6\%) \\
Unsure & 7 (2.5\%)\\
Other & 2 (0.7\%)  \\
\bottomrule  
\end{tabular}  
\label{tbl:communication_style}  
\end{table*} 

\section{Sample Responses from AI Productivity Agent}
\label{sec:sample_resposne_aiagent}
\begin{figure}[ht!]
     \begin{subfigure}[b]{0.75\textwidth}
\includegraphics[width=\textwidth]{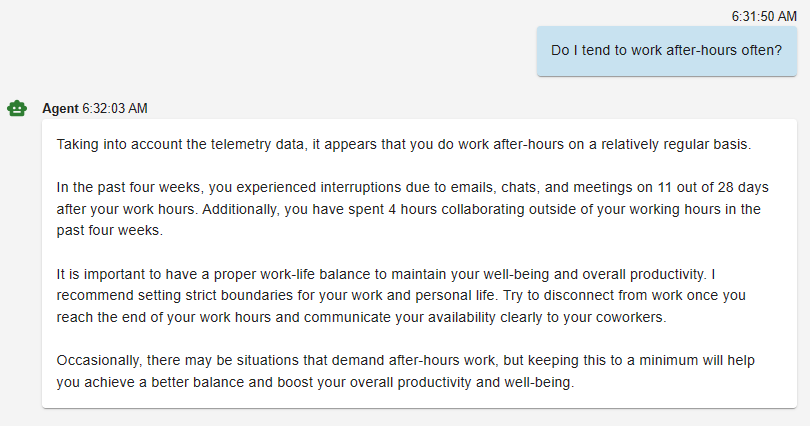}
        \caption{Sample Chat I}
        \label{fig:sample_chat2}
    \end{subfigure}
     \begin{subfigure}[b]{0.75\textwidth}
\includegraphics[width=\textwidth]{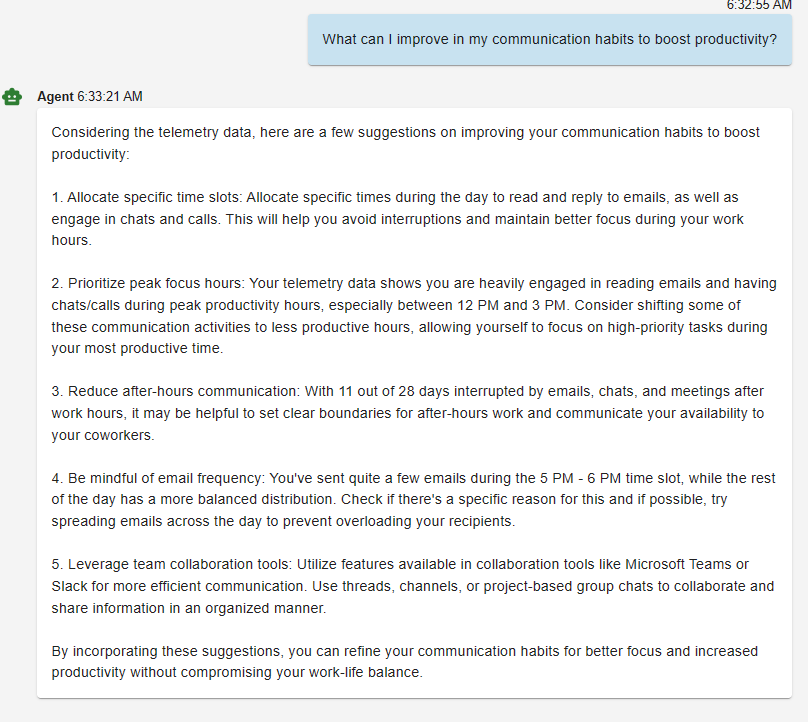}
        \caption{Sample Chat II}
        \label{fig:sample_chat3}
    \end{subfigure}
    \caption{Sample Responses}
\end{figure}

\section{Sample Personalized Narrative}

\begin{figure}[ht!]
     \begin{subfigure}[b]{0.75\textwidth}
\includegraphics[width=\textwidth]{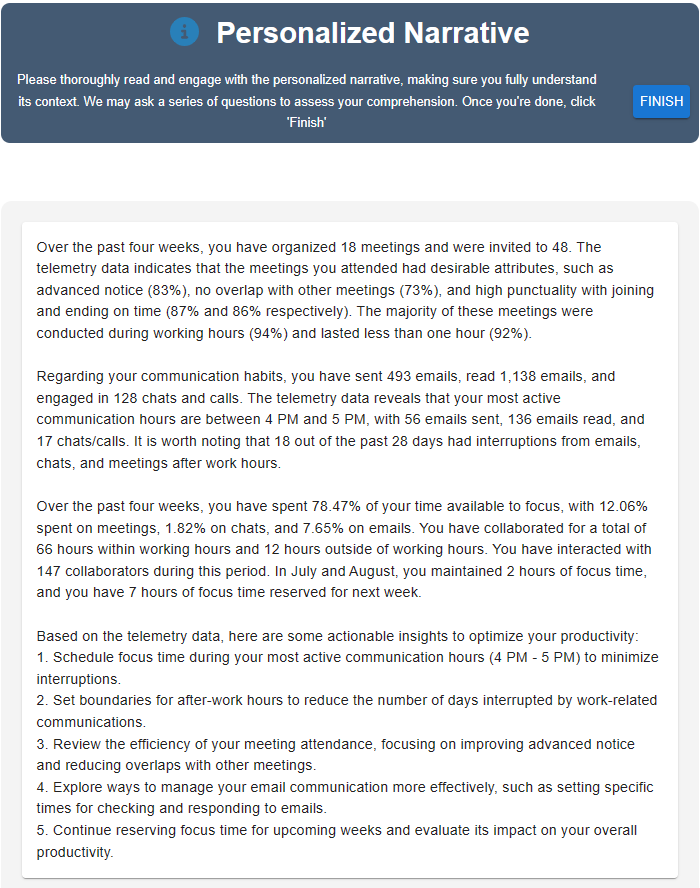}
        \label{fig:sample_narrative}
    \end{subfigure}
    \caption{Sample Personalized Narrative}
\end{figure}